\documentclass[%
 reprint,
 twocolumn,
 superscriptaddress,
 amsmath,amssymb,
 aps,
 pre,
]{revtex4-2}
 \usepackage[colorlinks]{hyperref}
\usepackage{graphicx}
\usepackage{dcolumn}
\usepackage{bm}
\usepackage[normalem]{ulem}
\usepackage{xcolor}
\usepackage{pdfpages} 
\usepackage{pgffor} 

\makeatletter
\AtBeginDocument{\let\LS@rot\@undefined}
\makeatother

\def\supplementfilename{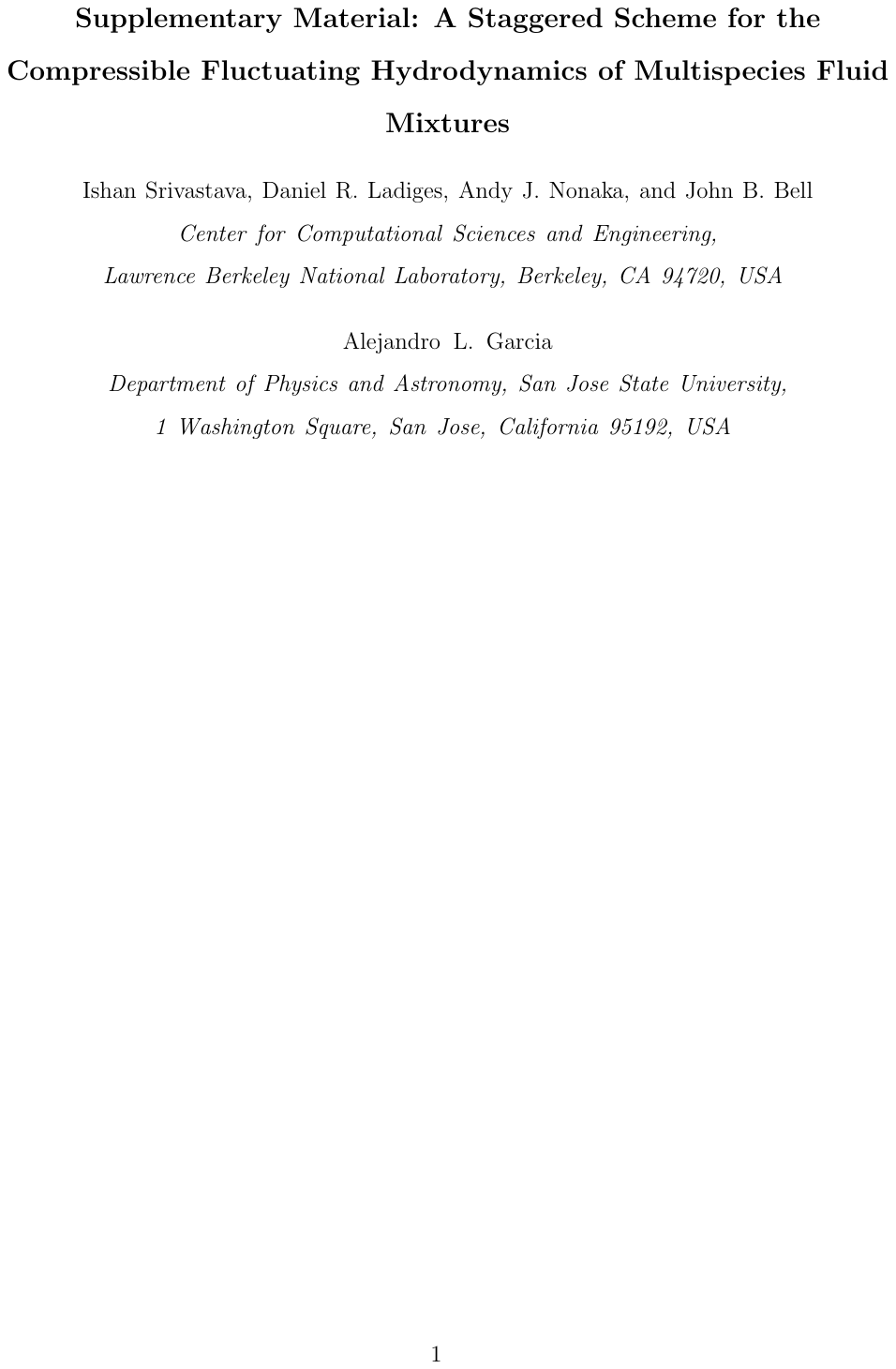}

\pdfximage{\supplementfilename}
\def\numbersupplementpages{\the\pdflastximagepages}

\newif\ifarXiv
\arXivtrue 

\newcommand{\half}{\frac{1}{2}}

\newcommand{\uFluid}[0]{\mathbf{v}}
\newcommand{\Momentum}[0]{\mathbf{j}}
\newcommand{\SpeciesFlux}[0]{{\boldsymbol{F}}}
\newcommand{\StressTensor}[0]{{\boldsymbol{\Pi}}}
\newcommand{\HeatFlux}[0]{{\boldsymbol{Q}}}

\newcommand{\nablaBold}[0]{\boldsymbol{\nabla}}

\newcommand{\Srivastava}[1]{{\bf[{\color{violet}#1}]}}

\newcommand{\commentout}[1]{}

\newcommand{\MarginPar}[1]   
{\marginpar{\vskip-\baselineskip 
\raggedright\tiny\sffamily\hrule\smallskip{\color{red}#1}\par\smallskip\hrule}}

\begin{document}


\title{A Staggered Scheme for the Compressible Fluctuating Hydrodynamics of Multispecies Fluid Mixtures}

\author{Ishan Srivastava}
    \email{isriva@lbl.gov}
    \affiliation{Center for Computational Sciences and Engineering, Lawrence Berkeley National Laboratory, Berkeley, California 94720, USA}
    
\author{Daniel R. Ladiges}
    \affiliation{Center for Computational Sciences and Engineering, Lawrence Berkeley National Laboratory, Berkeley, California 94720, USA}
    
\author{Andy J. Nonaka}
    \affiliation{Center for Computational Sciences and Engineering, Lawrence Berkeley National Laboratory, Berkeley, California 94720, USA}
    
\author{Alejandro L. Garcia}
    \affiliation{Department of Physics and Astronomy, San Jose State University, 1 Washington Square, San Jose, California 95192, USA}
    
\author{John B. Bell}
    \affiliation{Center for Computational Sciences and Engineering, Lawrence Berkeley National Laboratory, Berkeley, California 94720, USA}

\date{\today}

\begin{abstract}
We present a numerical formulation for the solution of non-isothermal, compressible, Navier-Stokes equations with thermal fluctuations to describe mesoscale transport phenomena in multispecies fluid mixtures. The novelty of our numerical method is the use of staggered grid momenta along with a finite volume discretization of the thermodynamic variables to solve the resulting stochastic partial differential equations. The key advantages of the numerical scheme are significantly simplified and compact discretization of the diffusive and stochastic momentum fluxes, and an unambiguous prescription of boundary conditions involving pressure. The staggered grid scheme more accurately reproduces the equilibrium static structure factor of hydrodynamic fluctuations in gas mixtures compared to a collocated scheme described previously in [\emph{Balakrishnan et al., Phys.~Rev.~E 89:013017, 2014}]. The numerical method is tested for ideal noble gases mixtures under various nonequilibrium conditions, such as applied thermal and concentration gradients, to assess the role of cross-diffusion effects, such as Soret and Dufour, on the long-ranged correlations of hydrodynamic fluctuations, which are also more accurately reproduced compared to the collocated scheme. We numerically study giant nonequilibrium fluctuations driven by concentration gradients, and fluctuation-driven Rayleigh-Taylor instability in gas mixtures. Wherever applicable, excellent agreement is observed with theory and measurements from the direct simulation Monte Carlo (DSMC) method.

\end{abstract}

\maketitle


\section{\label{sec:intro}Introduction}

The intrinsic thermal (Brownian) motion of the constituent atoms and molecules in a fluid at equilibrium results in hydrodynamic fluctuations that are well-understood from the results of classical thermodynamics~\cite{landau_statistical_1980}. When a fluid is maintained in a nonequilibrium state, such as under macroscopic gradients of temperature, chemical potential, or velocity such as in shear flow~\cite{garcia1987shear}, the influence of these hydrodynamic fluctuations can be enhanced by several orders of magnitude~\cite{Zara1,croccolo2016a}. Such enhancement occurs as a result of spatial coupling between the fluctuations of hydrodynamic variables, such as between temperature and velocity fluctuations in the presence of a thermal gradient~\cite{Mans1,Bell1}, and between concentration and velocity fluctuations in the presence of a concentration gradient~\cite{Done3}. These enhancements can also be induced by cross-diffusion phenomena such as thermodiffusion (Soret effect)~\cite{croccolo2012,ortizdezarate2014}. The long-ranged, scale-invariant structure of these correlated fluctuations is apparent from their structure factor that exhibits a power-law divergence\commentout{$S(k) \propto k^{-4}$, where $k$ is the wavenumber}~\cite{Zara1}. These ``giant fluctuations'' have been experimentally measured using shadowgraphy imaging of the diffusive interface of two miscible fluids in microgravity environments~\cite{Vailati1}, 
and their spatial extent is found to be only limited in size by either gravity~\cite{martinezpancorbo2017} or finite system size~\cite{giraudet2015}. The long-ranged nature of these hydrodynamic correlations introduces strong finite-size effects resulting in intriguing physical phenomena such as enhanced diffusive transport~\cite{Done3,Done4}, and fluctuation-induced Casimir-like forces~\cite{CasimirPrl2015} in nonequilibrium fluids. Hydrodynamic fluctuations 
also play a fundamental role in the vicinity of hydrodynamic instabilities such as
Rayleigh-Benard~\cite{bodenschatz2000recent}, Rayleigh-Taylor~\cite{kadau2004}, Richtmeyer-Meshkov~\cite{narayanan2018}, and Kelvin-Helmholtz~\cite{Bell2}. Furthermore, recent simulations have demonstrated that thermal fluctuations dominate the energy spectrum in the dissipation range of turbulence at length scales comparable to the Kolmogorov length~\cite{Gallis2022,bell_nonaka_garcia_eyink_2022}.

Particle-based numerical methods such as molecular dynamics (MD) and direct simulation Monte Carlo (DSMC) naturally capture the thermal fluctuations in fluids, but are computationally expensive even for coarse-grained methods such as the DSMC~\cite{LadigesMembrane19}. Specifically, the time scale of particle motion is significantly shorter than the time scale of hydrodynamic evolution, thus rendering many of these methods intractable for various problems of practical interest. Additionally, there are several outstanding challenges in simulating cross-diffusion phenomena using particle-based methods~\cite{ramirez-hinestrosa2021}, which are crucial in multispecies transport. 

Alternatively, thermal fluctuations can be introduced in a continuum framework by including a stochastic forcing in the Navier-Stokes conservation equations, as was originally proposed by Landau and Lifshitz~\cite{Land2}, and later extended to the case of fluid mixtures by Cohen et al.~\cite{CohenBinary1971}, and by Ottinger in the context of the GENERIC framework~\cite{ottinger2009}. In such a \emph{fluctuating hydrodynamics} (FHD) formulation, a stochastic flux is added to each dissipative flux associated with the transport of species mass, momentum and energy in a manner that satisfies the fluctuation-dissipation balance~\cite{ottinger2005beyond}. As such, the resulting system of stochastic partial differential equations (PDEs) ensures conservation at the microscopic scale while reproducing the Gibbs-Boltzmann distribution of thermal fluctuations in accordance with equilibrium statistical mechanics. The framework of FHD has been useful in understanding the behavior of fluids in various nonequilibrium conditions~\cite{Zara1,garrido2022}, but theoretical calculations have been feasible only with simplifying assumptions for boundary conditions and transport properties as well as ignoring the nonlinear coupling between hydrodynamic fields and cross-diffusion effects (such as the Dufour effect~\cite{Giov1}).

Various numerical methods have been proposed to solve FHD, starting with early works for a simple scheme to solve the stochastic heat equation~\cite{Garcia1987} and one-dimensional linearized fluctuating Navier-Stokes equation~\cite{Mans1}. Subsequently, finite-volume methods were developed, such as De Fabritiis et al.,~\cite{DeFabritiis2007} for compressible isothermal fluids, and a collocated finite-volume method for compressible, non-isothermal, binary fluid mixtures~\cite{Bell2}. Donev et al., improved these numerical schemes by using a specialized third-order Runge-Kutta (RK3) integrator that minimizes the temporal integration error in the discrete equilibrium structure factor of hydrodynamic fluctuations~\cite{Done1}. These methods for compressible FHD, originally formulated for single fluid species and binary mixtures, were extended to multispecies fluid mixtures, and were used to investigate giant fluctuations and Rayleigh-Taylor instability in ternary and four-species mixture respectively~\cite{Bala1}.

A staggered spatial discretization of the fluctuating Navier-Stokes equations provides several advantages compared to a collocated discretization. In the former, the fluid velocity and momenta reside on a grid that is staggered by half a finite-volume cell width to the grid that stores other conserved and thermodynamic fluid variables, such as density, pressure and temperature. A staggered-grid arrangement has distinct well-known advantages for incompressible flows~\cite{balboa2012,griffith2012}, but is also advantageous for compressible flows as demonstrated in ref.~\cite{balboa2012} for the FHD formulation of binary isothermal mixtures of gases. A staggered grid for the velocity results in a more compact stencil for the velocity update resulting from a pressure gradient, which provides for better handling of the pressure oscillations in numerical solutions. Because the velocity resides on the face of a centered finite-volume grid in the staggered scheme, no extra boundary conditions for pressure are needed for simulations with physical walls. Furthermore, for a spatially-varying viscosity, the complexities associated with discretizing the divergence of viscous flux (shear stress tensor)~\cite{Done1,Bala1} are considerably simplified when using staggered grids. In the context of FHD, a staggered grid also similarly prescribes a simpler, compact discretization of the stochastic momentum fluxes~\cite{balboa2012}. The advantages of a staggered grid formulation for FHD have been discussed previously for isothermal compressible flows for single species~\cite{delgado-buscalioni2008,voulgarakis2009a} and binary fluid mixtures~\cite{balboa2012}. 

In this paper, we describe a staggered grid formulation for the FHD of multispecies fluid mixtures using compressible Navier-Stokes equations without the isothermal assumption. We demonstrate that the staggered method more accurately reproduces the equilibrium static structure factor for multispecies fluids, which is the steady-state covariance of the fluctuating fields in the Fourier space. We use our numerical scheme to calculate hydrodynamic correlations for fluid mixtures in various nonequilibrium scenarios. We find excellent correspondence with DSMC simulations and obtain more accurate predictions compared to a cell-centered scheme for compressible, multispecies, non-isothermal equations of FHD~\cite{Bala1}. Particularly, we measure giant fluctuations in simulations of binary mixtures of ideal gases under external concentration gradients and observe a power-law divergence of the structure factor of hydrodynamic fluctuations at low wavenumbers. We also investigate the role of cross-diffusive effects, such as the combined Soret and Dufour effects, in the long-ranged correlations of hydrodynamic fluctuations in ideal gas mixtures that are typically ignored in theoretical FHD analyses. Lastly, we demonstrate the capability of our numerical method to model hydrodynamic instabilities driven only by thermal fluctuations by simulating the Rayleigh-Taylor instability in a fluid mixture where an initially perfectly flat interface separates the heavier and lighter fluid mixtures.

The paper is organized as follows. The mathematical theory of compressible FHD in multispecies mixtures is reviewed in Section~\ref{sec:theory}, and the staggered numerical method is described in Section~\ref{sec:numeric}. Computational results for equilibrium and nonequilibrium steady states of multispecies fluid mixtures are presented in Section~\ref{sec:results}, along with an emphasis on validation by DSMC simulations wherever applicable. Conclusions and avenues for future research are provided in Section~\ref{sec:conclusions}.

\section{\label{sec:theory}Mathematical Theory}
In this section we summarize the hydrodynamic equations that describe the FHD of multispecies, non-reacting fluid mixtures, and refer the reader to Balakrishnan et al.~\cite{Bala1} for a detailed treatment~\footnote{See Supplemental Material for: (a) typographical errors in ref.~\cite{Bala1} along with the corrections; (b) choices of numerical discretization of face-based advective fluxes of momentum. Comparisons of the equilibrium structure factors of gas mixtures corresponding to results in Sec.~\ref{sec:resultsA} and spatial correlations of hydrodynamic fluctuations in a binary gas mixture under thermal gradient corresponding to results in Sec.~\ref{sec:resultsC} are also provided for the various choices of face-based advective fluxes; (c) the discrete representation of the viscous stress tensor in the staggered-grid scheme; (d) comparisons of the mean profiles of hydrodynamic quantities and the correlations of hydrodynamic fluctuations between simulations that include and do not include Soret and Dufour effects.}.
We use a standard Fickian treatment based on the Maxwell-Stefan equations to formulate multispecies diffusion including cross-transport phenomena such as the Dufour and Soret effects; see, for example, Giovangigli~\cite{Giov1} . Although our formulation is general, we will study the specific case of ideal gas mixtures, whose transport coefficients are modeled based on the prescription by Giovangigli~\cite{Giov1}. 

\subsection{\label{subsec:hydro}Multispecies fluctuating hydrodynamics}
The equations of FHD are obtained by adding white Gaussian noise to each dissipative flux responsible for the entropy production in the deterministic compressible Navier-Stokes equations for multispecies mixtures~\cite{Giov1}. These stochastic fluxes represent thermal fluctuations in the fluid. Consider a multispecies fluid mixture with $N_s$ non-reacting species. The partial density of a
species $k$ is $\rho_k$, and the total density $\rho = \sum_k \rho_k$ is obtained by summing over all the partial densities. The mass fraction for species $k$ is $Y_k = \rho_k/\rho$, and $X_k = n_k/\sum_i n_i$ is the mole fraction where $n_i$ is the species number density. The mass fraction and mole fraction are related by $Y_k = (m_k/\overline{m})X_k$, where $m_k$ is the species molecular mass, and $\overline{m}=\left(\sum_{k} Y_k/m_k\right)^{-1}$ is the average molecular mass of the mixture.
The fluctuating Navier-Stokes equations of continuity, species, momentum and energy in the conservation form are given as~\cite{Espa1,Bala1}:
\begin{widetext}
\begin{eqnarray}
&\frac{\partial}{\partial t} \left( \rho \right) = - \nablaBold \cdot \left( \rho \uFluid \right), \label{eqn:cont} \\
&\frac{\partial }{\partial t} \left( \rho_k \right) = - \nablaBold \cdot \left( \rho_k \uFluid \right) 
- \nablaBold \cdot \left[ \SpeciesFlux_k + \widetilde{\SpeciesFlux}_k \right], \label{eqn:species} \\
&\frac{\partial }{\partial t} \left( \rho \uFluid \right) =  - \nablaBold \cdot \left[ \rho {\uFluid \otimes \uFluid + p\mathbb{I} } \right] - \nablaBold \cdot \left[ \StressTensor + \widetilde{\StressTensor} \right] + \rho \mathbf{g}, \label{eqn:mom} \\
&\frac{\partial }{\partial t} \left( \rho E \right) = - \nablaBold \cdot \left[\uFluid \left(\rho E + p\right) \right] 
- \nablaBold \cdot \left[ \HeatFlux + \widetilde{\HeatFlux} \right] -\nablaBold \cdot \left[ \left( \StressTensor + \widetilde{\StressTensor} \right) \cdot  \uFluid \right] + \rho \mathbf{g} \cdot \uFluid , \label{eqn:energy}
\end{eqnarray}
\end{widetext}
where $\mathbb{I}$ is the identity tensor, $\uFluid$ is the fluid velocity vector, $p$ is the pressure, $T$ is temperature, and $\rho \mathbf{g}$ is the vector of external body force density such as due to gravity. The total energy density $\rho E = \rho e + \frac12 \rho (\uFluid \cdot \uFluid)$ is the sum of internal energy and kinetic energy, where $e$ is the specific internal energy not including the gravitational potential energy. The tensor $\uFluid \otimes \uFluid$ is the outer product of $\uFluid$ with itself. The diffusive fluxes of species density $\SpeciesFlux_k$, diffusive momentum flux given by the viscous stress tensor $\StressTensor$, and the diffusive heat flux $\HeatFlux$ are governed by the transport properties of the fluid. In the FHD framework, the deterministic Navier-Stokes equations are augmented with stochastic fluxes that are denoted by the terms appearing with a tilde in Eqs.~\ref{eqn:species} -~\ref{eqn:energy}. These stochastic fluxes are formulated using the fluctuation-dissipation relations to produce thermodynamically-consistent covariances of the hydrodynamic fields at equilibrium~\cite{Land2,Zara1}.

The complete hydrodynamic description of a fluid based on the conservation Eqs.~\ref{eqn:cont} -~\ref{eqn:energy} requires the knowledge of 
properties of the fluid, such as the equation of state and specific heats. 
Although the formulation given above is quite general, here we will restrict our consideration to inert ideal gas mixtures.
For such a system, the equation of state is
\begin{equation}
    p = \frac{\rho k_B T}{\overline{m}}.
\end{equation}
where $k_B$ is Boltzmann's constant. The specific internal energy $e$ and enthalpy $h$ of the mixture are then defined as 
\begin{eqnarray}
    e(T,Y_k) &=& \sum_{k}Y_k e_k(T), \nonumber \\
    h(T,Y_k) &=& \sum_{k}Y_k h_k(T),
\end{eqnarray}
where $e_k$ and $h_k$ are partial specific internal energies and enthalpies respectively.  In an ideal gas, $e_k$ and $h_k$ are functions of $T$ only.  The species enthalpy and energy are related by
\begin{equation}
    h_k = e_k + \frac{k_B}{m_k}T.
\end{equation}
%
The specific heats at constant volume and pressure for the mixture are, respectively,
\begin{eqnarray}
    c_v(T) = \left(\frac{\partial e}{\partial T}\right)_{Y_k,v} &=& \sum_k Y_k c_{v,k}(T), \nonumber \\
    c_p(T)  = \left(\frac{\partial h}{\partial T}\right)_{Y_k,p} &=& \sum_k Y_k c_{p,k}(T).
\end{eqnarray}
Generally, the specific heats of each species are specified as a function of $T$, and the species energy and enthalpy are determined by an integration of the specific heats. Here we will consider the case of calorically perfect gases with constant specific heats.

The chemical potential per unit mass for each species in an ideal gas mixture is expressed as~\cite{Giov1}
\begin{equation}
    \mu_k = \frac{k_B T}{m_k}\left(\mathrm{ln}\; X_k + \mathrm{ln}\; \frac{p}{p_\text{o}} \right) + \mu_{k,\text{o}}(T),
\end{equation}
where $\mu_{k,\text{o}}(T)$ is the chemical potential of pure species $k$ at a 
reference pressure $p_\text{o}$.

\subsection{\label{subsec:viscosity}Viscous fluxes}

For the Newtonian fluids considered here, the components of the viscous stress tensor are
\begin{equation}
\Pi_{ij} = -\eta \left( \frac{\partial u_i}{\partial x_j} + \frac{\partial u_j}{\partial x_i}  \right) - \delta_{ij} \left( ( \kappa - \frac{2}{3} \eta ) {\bf \nabla} \cdot {\uFluid} \right),
\label{eqn:stress}
\end{equation}
where $\delta_{ij}$ is the Kronecker delta, $\eta$ is the shear viscosity, and $\kappa$ is the bulk viscosity. The viscosities, as well as the other transport coefficients, are not treated as constants but depend on the local state of the gas mixture~\cite{Giov1}.  

The stochastic stress needed to produce the thermodynamically correct velocity covariance is a Gaussian random field $\widetilde{\StressTensor}$, with zero ensemble mean $\langle \widetilde{\StressTensor} \rangle = 0$, and the following covariance \cite{Land2,Zara1}:
\begin{widetext}
\begin{eqnarray}
    \langle \widetilde{\Pi}_{ij}(\mathbf{r},t)\; , \;\widetilde{\Pi}_{mn}(\mathbf{r}',t')\rangle &=& \delta\left(\mathbf{r}-\mathbf{r}'\right) \delta\left(t-t'\right) [2k_B T \eta \left(\delta_{im}\delta_{jn}+\delta_{in}\delta_{jm}\right) \nonumber \\ 
    &+& 2k_B T\left(\kappa - \frac{2}{3}\eta\right)\delta_{ij}\delta_{mn}],
    \label{stochstress:corr}
\end{eqnarray}
\end{widetext}
where $\delta\left(\mathbf{r}-\mathbf{r}'\right)$ and $\delta\left(t-t'\right)$ are Dirac delta functions. Espa\~{n}ol~\cite{Espa1} described a following efficient form of the stochastic stress tensor $\widetilde{\StressTensor}$ in a three-dimensional system that produces the correct covariance:
\begin{equation}
    \widetilde{\StressTensor}(\mathbf{r},t) = \sqrt{2k_B T \eta} \widetilde{\mathcal{Z}} +
\left( \sqrt{\frac{k_B \kappa T}{3}} - \frac{\sqrt{2k_B \eta T}}{3} \right ) \mathrm{Tr} ( \widetilde{\mathcal{Z}} ) \mathbb{I},
\label{eq:stochstress}
\end{equation}
where
\begin{equation}
    \widetilde{\mathcal{Z}} = \frac{1}{\sqrt{2}}\left(\mathcal{Z}+\mathcal{Z}^{T}\right).
    \label{stoch:Z}
\end{equation}
is a symmetric matrix constructed from an uncorrelated Gaussian tensor field $\mathcal{Z}$ with zero mean and unit variance.

We note that the stochastic stress term defined by Eq. (\ref{eq:stochstress}) is not correct for quasi-2D and quasi-1D versions of FHD where no transport occurs in one and two directions respectively. In a finite-volume sense, these versions are equivalent to a system domain that is only one cell wide in one direction for quasi-2D simulations, and one cell wide in two directions for quasi-1D simulations.
Because we describe numerical simulations in this paper that include results from both quasi-1D and quasi-2D simulations in addition to full 3D simulations, the modifications required for computing the components of $\widetilde{\StressTensor}$ in lower dimensions are described in Appendix~\ref{AppA}.

\subsection{\label{subsec:species}Multispecies diffusion and heat flux}
The classical construction of multispecies diffusion was provided by de Groot and Mazur~\cite{de1962non} through the Onsager form for $N_s-1$ species fluxes, where the diffusive flux for the last species is determined by the continuity equation. However, such a construction requires the identification of a distinguishable reference species, which can cause numerical issues in simulations where the reference species is only present in trace amounts. A numerically stable, full system construction for $N_s$ species fluxes was derived in Balakrishnan et al.,~\cite{Bala1}, where it was demonstrated that its entropy production is exactly identical to the Onsager form. In this work we utilize such a full construction of multispecies diffusion, and remark that a similar derivation was given by Ottinger using the GENERIC formalism~\cite{ottinger2009}. We use Einstein indices in this subsection to describe the various fluxes, where the Greek symbols in the subscripts denote the species index, and $i$ and $j$ subscripts represent the spatial directions $x,y$ and $z$.

The $i$-th spatial component of the flux of species $\alpha$, $\SpeciesFlux_{\alpha i}$, is expressed as
\begin{equation}
    \SpeciesFlux_{\alpha i} = -\frac{1}{T}\mathbf{L}_{\alpha \beta}\left[\mu_{\beta,i_{T}} + \frac{\xi_{\beta}}{T}T_{,i} \right],
\label{specflux}
\end{equation}
where $\mathbf{L}$ is an $N_s\times N_s$ augmented Onsager matrix that depends on the multispecies diffusion coefficients, $\mu_{\beta,i_{T}}$ is the $i$-th spatial derivative of the chemical potential $\mu_{\beta}$ of species $\beta$ at constant $T$, and $\xi_{\beta}$ is related to the rescaled thermal diffusion (Soret) coefficient of species $\beta$, as specified below. The deterministic heat flux is given by
\begin{equation}
    \HeatFlux_{i} = -\zeta \frac{T_{,i}}{T^2} + \left(\xi_{\alpha} + \mathbf{h}_{\alpha}\right)\SpeciesFlux_{\alpha i},
\label{heateq}
\end{equation}
where $\mathbf{h}_{\alpha}$ is the enthalpy of species $\alpha$, $\zeta$ depends on the multispecies mixture thermal conductivity, and $T_{,i}$ is the $i$-th spatial derivative of $T$.

Although useful when determining the stochastic fluxes, the Onsager form of the species diffusion given by Eq.~\ref{specflux} is typically not used in numerical simulations because of the singularities in the gradient of chemical potential as a species vanishes, i.e., $X_k \to 0$. This issue is avoided by recasting $\SpeciesFlux$ in a Fickian form given by 
\begin{equation}
    \SpeciesFlux_{\alpha i} = \rho \boldsymbol{\mathcal{Y}}_{\alpha} \boldsymbol{\mathcal{D}}_{\alpha\beta} \left( \mathbf{X}_{\beta,i} +  \frac{(\mathbf{X}_{\beta}-\mathbf{Y}_{\beta})}{p} p_{,i} + \frac{\boldsymbol{\mathcal{X}}_{\beta} \widetilde{\chi}_{\beta} }{T} T_{,i}  \right),
    \label{eq:specflux_ideal}
\end{equation}
where $\boldsymbol{\mathcal{D}}$ is the $N_s\times N_s$ multispecies diffusion matrix computed from the binary diffusion coefficients using Maxwell-Stefan relations (see Giovangigli~\cite{Giov1} for additional details.) Here the column vectors of species mole fractions $X_k$ and mass fractions $Y_k$ are represented by $\mathbf{X}$ and $\mathbf{Y}$ respectively, whereas $\boldsymbol{\mathcal{Y}}$ and $\boldsymbol{\mathcal{X}}$ are $N_s\times N_s$ diagonal matrices of $Y_k$ and $X_k$ respectively. The strength of thermal diffusion (Soret effect) is governed by the $N_s\times 1$ column vector of rescaled thermal diffusion ratios $\widetilde{\chi}$. The barodiffusion term associated with the spatial derivative of pressure $p_{,i}$ is a thermodynamic effect and it does not have an associated transport coefficient \cite{Land2}. The matrix  $\boldsymbol{\mathcal{D}}$ is symmetric and positive semidefinite with $\boldsymbol{Y}$ being in the null space of $\boldsymbol{\mathcal{D}}$.
This property ensures that that $\sum_{\alpha}\SpeciesFlux_{\alpha} = 0$.

The deterministic heat flux $\HeatFlux$ is also recast in terms of the mixture thermal conductivity $\lambda$ and rescaled thermal diffusion ratios $\widetilde{\chi}$, which are relatively inexpensive to compute, to obtain
\begin{equation}
    \HeatFlux_{i} = -\lambda T_{,i} + \left(k_B T \widetilde{\chi}_{\alpha} \boldsymbol{\mathcal{M}}^{-1}_{\alpha} + \mathbf{h}_{\alpha}\right) \SpeciesFlux_{\alpha i},
    \label{heat_ideal}
\end{equation}
where $\boldsymbol{\mathcal{M}}$ is a diagonal matrix of the molecular masses of the species.

The stochastic counterparts of the deterministic diffusion and heat fluxes are formulated by considering that the stochastic fluxes are uncorrelated in space and time, and can be represented as white noises, such as:
\begin{eqnarray}
    \widetilde{\SpeciesFlux}_{i} &=& \mathbf{B}\mathcal{Z}^{(\SpeciesFlux;i)}, \nonumber \\
    \widetilde{\HeatFlux}_{i} &=& \sqrt{\zeta}\mathcal{Z}^{(\HeatFlux;i)} + \left(\xi^{T}+\mathbf{h}^{T}\right)\widetilde{\SpeciesFlux}_{i}.
    \label{eq:stochfluxes}
\end{eqnarray}
where $i = x,y,z$ is the spatial direction, and $\mathcal{Z}^{(\HeatFlux;i)}$ is a scalar and $\mathcal{Z}^{(\SpeciesFlux;i)}$ is a vector of $N_s$ independent Gaussian white-noise terms such that
\begin{eqnarray}
\langle \mathcal{Z}^{(\SpeciesFlux_\alpha;i)}(\mathbf{r},t) \mathcal{Z}^{(\SpeciesFlux_\beta;j)}(\mathbf{r}',t') \rangle &=& \delta_{ij}\delta_{\alpha\beta}\delta(\mathbf{r}-\mathbf{r}')\delta(t-t') \nonumber \\ 
\langle \mathcal{Z}^{(\HeatFlux;i)}(\mathbf{r},t) \mathcal{Z}^{(\HeatFlux;j)}(\mathbf{r}',t')\rangle &=& \delta_{ij}\delta(\mathbf{r}-\mathbf{r}')\delta(t-t') \nonumber \\
\langle \mathcal{Z}^{(\HeatFlux;i)} \mathcal{Z}^{(\SpeciesFlux;j)} \rangle &=& 0,
\end{eqnarray}
where $\alpha$ and $\beta$ above correspond to species indices, and $i,j = x,y,z$ are the spatial directions. To satisfy fluctuation-dissipation balance, the amplitude of the noise is set such that $\mathbf{B}\mathbf{B}^{T} = 2k_B\mathbf{L}$~\cite{Bala1}. The matrix $\mathbf{B}$ is not uniquely defined, and numerically a lower-triangular $\mathbf{B}$ can be computed from the Cholesky factorization of $\mathbf{L}$. We note that the properties of $\mathbf{L}$ 
ensures that that $\sum_{\alpha}\widetilde{\SpeciesFlux}_{\alpha} = 0$~\cite{Bala1}.

Comparing the two representations, it is straightforward to show that
\begin{eqnarray}
    \mathbf{L} &=& \frac{\rho \bar{m}}{k_B}\boldsymbol{\mathcal{Y}} \boldsymbol{\mathcal{D}} \boldsymbol{\mathcal{Y}}, \nonumber \\
    \zeta &=& T^2 \lambda, \nonumber \\
    \xi &=& k_B T \boldsymbol{\mathcal{M}}^{-1} \widetilde{\chi}.
\end{eqnarray}
A full specification of the transport coefficients requires the computation of $\boldsymbol{\mathcal{D}}$, $\lambda$ and $\widetilde{\chi}$.
Efficient techniques for computing these coefficients are discussed in Ern and Giovangigli~\cite{ern1994multicomponent} and Giovangigli~\cite{Giov1}.

In this paper we have restricted consideration to noble gases modeled using the hard sphere properties; however, the methodology can be easily extended to more general ideal mixtures 
by utilizing existing software for computing transport properties such as the EGLIB package that is commonly used in the reactive flow community~\cite{ern2004eglib}. In addition, the framework described above can be extended to include chemical reactions~\cite{Bhat2}, and multiphase flows such as previously demonstrated for multiphase flow of a single species of a van der Waals fluids near the critical point~\cite{Chau1}. The numerical framework can also be extended to model non-ideal fluids by incorporating additional considerations associated with the coarse-graining of thermal fluctuations at the hydrodynamic scales~\cite{Parsa_Wagner_2020}.

\section{\label{sec:numeric}Numerical Method}
In this section we describe our numerical method to solve the fluctuating Navier-Stokes (FNS) equations described in Eqs.~\ref{eqn:cont}-~\ref{eqn:energy}. The current method extends a previous staggered grid scheme for isothermal, compressible FNS equations for binary mixtures~\cite{balboa2012} to incorporate non-isothermal effects and to treat multispecies mixtures. Therefore, the Dufour cross-diffusion effect that describes energy transport due to species diffusion is naturally simulated in our scheme, and which was absent in the isothermal formulation. We will focus the description of our scheme on spatial discretization of the fluctuating energy conservation equation, Eq.~\ref{eqn:energy}, which is required to capture the non-isothermal effects. Compared to the isothermal version, the new algorithm uses state-dependent transport coefficients and models the Soret effect dynamically rather than a fixed external force. The new staggered algorithm represents an alternative to a previous collocated finite-volume scheme for compressible, non-isothermal FNS equations for multispecies gas mixtures~\cite{Bala1}. In Sec. \ref{sec:results} we demonstrate that the present staggered grid scheme more accurately reproduces the hydrodynamic fluctuations in both equilibrium and nonequilibrium settings.

A staggered-grid approach based on the method-of-lines approach is used to spatially discretize the system of stochastic PDEs, and the resulting stochastic ordinary differential equations are integrated explicitly in time using a low-storage third-order Runge-Kutta (RK3) integrator~\cite{Done1,Bala1}. The FNS equations can be represented in the following compact form:
\begin{equation}
    \partial_{t}\mathbf{U} = -\nabla\cdot\mathbf{F}_{H} -\nabla\cdot\mathbf{F}_{D} -\nabla\cdot\mathbf{F}_{S} + \mathbf{H} \equiv \mathbf{R}\left(\mathbf{U},Z\right),
    \label{FHD}
\end{equation}
where $\mathbf{U}$ represents the set of conserved hydrodynamic fields:
\begin{equation}
    \mathbf{U} = 
    \begin{bmatrix}
    \rho \\ \rho \mathbf{Y}\\ \rho \uFluid \\ \rho E
    \end{bmatrix},
\label{compact}
\end{equation}
$\mathbf{F}_{H}$, $\mathbf{F}_{D}$ and $\mathbf{F}_{S}$ are the hyperbolic, diffusive and stochastic fluxes respectively, and $\mathbf{H}$ is a term representing external forcing. The right hand side of Eq.~\ref{FHD} is represented by $\mathbf{R}$, and $Z$ is the spatiotemporal discretization of the Gaussian random field $\mathcal{Z}$ that represents the noise used to construct the stochastic fluxes. The various fluxes are:
\begin{eqnarray}
    \mathbf{F}_{H} &=& 
    \begin{bmatrix}
    \rho \uFluid \\ \rho \uFluid \mathbf{Y} \\ \rho \uFluid \uFluid^{T} + p\mathbf{I} \\ \uFluid(\rho E + p)
    \end{bmatrix};
    \qquad
    \mathbf{F}_{D} =
    \begin{bmatrix}
    0 \\ \SpeciesFlux \\ \StressTensor \\ \HeatFlux + \StressTensor\cdot\uFluid
    \end{bmatrix}; \\
    \mathbf{F}_{S} &=&
    \begin{bmatrix}
    0 \\ \widetilde{\SpeciesFlux} \\ \widetilde{\StressTensor} \\ \widetilde{\HeatFlux} + \widetilde{\StressTensor}\cdot\uFluid
    \end{bmatrix}.
\end{eqnarray}
The external forcing considered here is of the form $\mathbf{H}=\left[0,0,\rho \mathbf{g},\rho \mathbf{g}\cdot\uFluid\right]^{T}$, representing forcing due to gravitational acceleration $\mathbf{g}$, but it can generally represent a more complex external forcing.

\begin{figure*}[t]
\centering
\includegraphics{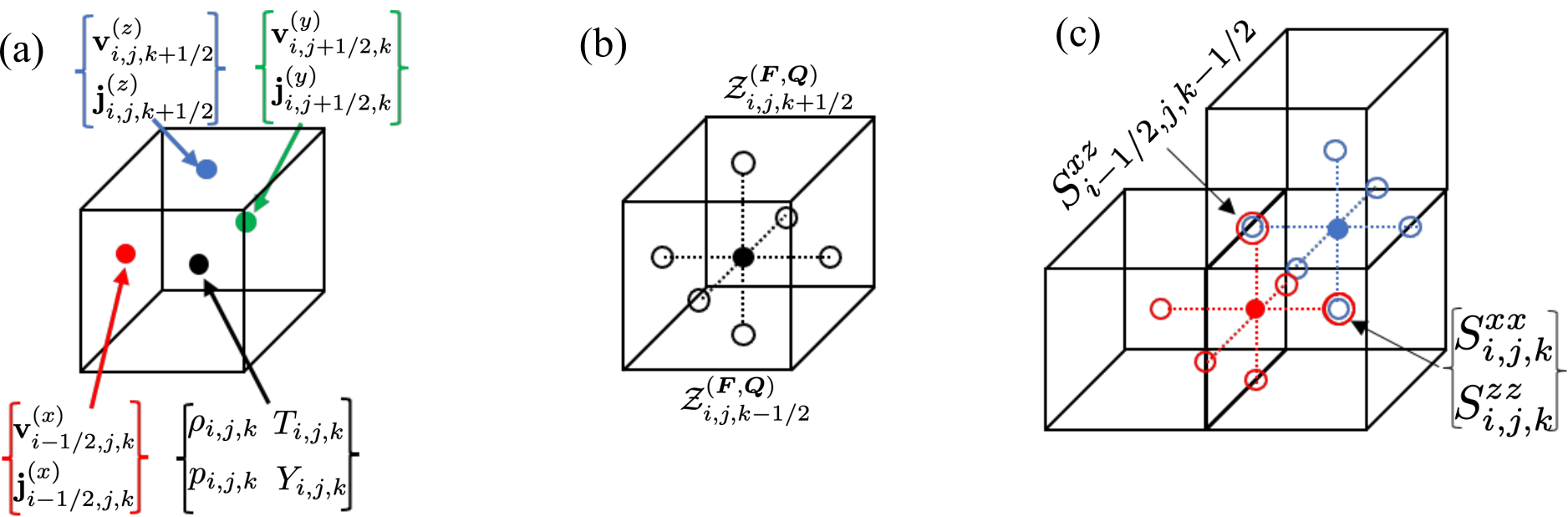}
\caption{An illustration of the staggered spatial discretization. (a) The thermodynamic variables such as $p$, $\rho$, $T$ and $\mathbf{Y}$ are located at the cell centers, and components of vector quantities such as $\uFluid^{(x)}$ and $\mathbf{j}^{(x)}$ (red), $\uFluid^{(y)}$ and $\mathbf{j}^{(y)}$ (green), and $\uFluid^{(z)}$ and $\mathbf{j}^{(z)}$ (blue) are located on respective staggered grids, i.e., on the faces of the finite volume grid. (b) Random numbers $\mathcal{Z}^{(\SpeciesFlux,\HeatFlux)}$ corresponding to $\widetilde{\SpeciesFlux}$ and $\widetilde{\HeatFlux}$ are generated at the faces of the control volume centered around the location of the thermodynamic variables. (c) The random numbers $S^{xx}$ and $S^{zz}$ corresponding to the diagonal terms of $\widetilde{\StressTensor}$ are generated at the faces of the staggered control volumes around $\mathbf{j}^{(x)}$ (red) and $\mathbf{j}^{(z)}$ (blue) respectively that are are collocated at the cell centers of the finite volume grid. The random numbers $S^{xz}$ corresponding to the off-diagonal terms of $\widetilde{\StressTensor}$ are generated on faces of the staggered control volumes around both $\mathbf{j}^{(x)}$ and $\mathbf{j}^{(z)}$ that are collocated on the edges of the finite volume grid. The dotted lines in (b) and (c) represent the six-point divergence operator.}
\label{fig0}
\end{figure*}

The numerical method described here is implemented within the AMReX framework~\cite{zhang2019a}, which uses an MPI paradigm for massively-parallel simulations along with GPU-based performance acceleration. The code for the staggered FHD numerical method has been tested on massively-parallel multicore architectures, and it is compatible with several GPU programming environments where significant performance enhancement by up to a factor of $20$ has been observed upon using GPU accelerators with near perfect scaling because of the explicit time integration scheme. The numerical method has been implemented in our fluctuating hydrodynamics software, FHDeX, and it is available online as an open-source code \footnote{FHDeX website, \url{https://github.com/AMReX-FHD/FHDeX}}.

\subsection{\label{subsec:spatial}Spatial discretization}
The equations of compressible FHD are spatially discretized on a uniform Cartesian grid with spacing given by $\Delta x$, $\Delta y$ and $\Delta z$ in the $x$, $y$ and $z$ directions respectively.  In a staggered grid spatial discretization, the conserved scalar variables $\rho$, $\rho Y_k$ and $\rho E$, and primitive scalar variables $p$, $T$, $X_k$ and $Y_k$ are discretized at the centers of a cell ($i,j,k$), whereas the vector variables, such as conserved momentum density $\Momentum = \rho \uFluid$ and velocity $\uFluid$ are discretized on the normal faces of the grid, as shown in Fig.~\ref{fig0}(a). As such, the $x$ component of velocity $\uFluid^{(x)}$ and momentum density $\Momentum^{(x)}$ reside at point $(i+\frac{1}{2},j,k)$ of the cell face, the $y$ component of velocity $\uFluid^{(y)}$ and momentum density $\Momentum^{(y)}$ reside at point $(i,j+\frac{1}{2},k)$ of the cell face, and so on. Regardless of where particular variables are defined, we consider them as representing fluctuating quantities over a cell of volume $\Delta x \Delta y \Delta z$. This type of staggered discretization is commonly used in projection algorithms for incompressible flows~\cite{balboa2012}, and it was also used previously to simulate isothermal compressible flows~\cite{balboa2012}.  Here we demonstrate its usage for the first time for non-isothermal, multispecies compressible equations of FHD.

The computation of cell-centered primitive scalar variables $p$ and $T$ from the conserved variables through the equation of state requires computing a cell-centered kinetic energy density $K$ from the momentum densities located at the faces of the grid. This is achieved by a simple face to center interpolation such as:
\begin{widetext}
\begin{equation}
    K_{i,j,k} = \frac{1}{8\rho_{i,j,k}}\left[\left(\Momentum^{(x)}_{i-\half,j,k}+\Momentum^{(x)}_{i+\half,j,k}\right)^{2} + \left(\Momentum^{(y)}_{i,j-\half,k}+\Momentum^{(y)}_{i,j+\half,k}\right)^{2} + \left(\Momentum^{(z)}_{i,j,k-\half}+\Momentum^{(z)}_{i,j,k+\half}\right)^{2}\right].
\end{equation}
\end{widetext}
Once $K$ is determined, the internal energy $e$ is computed, which is used to determine the temperature $T$ and pressure $p$ of the gas mixture. Upon specifying all the primitive thermodynamic variables, the transport coefficients are evaluated at the cell centers.

The advective fluxes for the total density, species density and energy density are computed at the faces of the grid, corresponding to the normal directional component of each flux. The staggered velocities on the faces required for computing the advective fluxes are defined by interpolating the density from the cell centers to the faces such as:
\begin{equation}
    \uFluid^{(x)}_{i+\frac{1}{2},j,k} = \frac{2\Momentum^{(x)}_{i+\frac{1}{2},j,k}}{\rho_{i,j,k} + \rho_{i+1,j,k}}.
\end{equation}
Therefore, the advective flux $\rho \uFluid$ of the density is simply the momentum density $\Momentum$ on the faces of the grid. For every other cell-centered scalar conserved quantity $\phi$ such as species density $\rho \mathbf{Y}$ and energy density $\rho E$, the advective fluxes are defined on the face by a simple interpolation the conserved quantities as:
\begin{equation}
    \left(\phi \uFluid\right)^{(x)}_{i+\frac{1}{2},j,k} = \frac{1}{2}\left(\phi_{i+1,j,k} + \phi_{i,j,k}\right)\uFluid^{(x)}_{i+\frac{1}{2},j,k},
    \label{eq:advec_methodA}
\end{equation}
although higher-order interpolation schemes such as the piecewise parabolic method can also be used~\cite{colella1984}.The advective fluxes given by Eq.~\ref{eq:advec_methodA} were previously demonstrated to satisfy the discrete fluctuation-dissipation balance in isothermal, compressible FHD equations~\cite{balboa2012}. The face-based advective fluxes can alternatively be defined using momentum on the staggered grid along with an interpolation of the thermodynamic variables from the cell centers to the faces. We have verified that these choices for discretizing face-based advective fluxes only minimally influence the hydrodynamic fluctuations in equilibrium and nonequilibrium conditions, and we refer the reader to the Supplemental Material for further numerical details and comparisons~\cite{Note1}. The results presented in the paper use the definition in Eq.~\ref{eq:advec_methodA} to discretize the advective fluxes. The cell-centered divergence of these advective fluxes on the faces is computed by the standard discrete six-point divergence operator $\mathbf{D}^{f\to c}$, as shown in Fig.~\ref{fig0}(b). It was demonstrated that such a discrete divergence operator for the advective fluxes in a staggered grid is skew-symmetric~\cite{balboa2012}, which is important to satisfy the discrete fluctuation-dissipation balance, i.e., advection should not modulate the fluctuations but only transport them~\cite{Done1}.

The advection of the momentum density vector is a bit more involved than the scalar quantities. The advective flux for the component of the momentum normal to a face is stored at the cell centers of the finite-volume grid, whereas the advective flux for the tangential momentum is stored at the edges of the grid. For example, the discrete advective flux $\left(\uFluid^{(x)}\Momentum^{(x)} + p\right)$ is computed by a simple interpolation of face-centered velocity and momentum to cell centers as
\begin{widetext}
\begin{equation}
    \left(\uFluid^{(x)}\Momentum^{(x)} + p\right)_{i,j,k} = \frac{1}{4}\left[\left(\uFluid^{(x)}_{i+\frac{1}{2},j,k}+\uFluid^{(x)}_{i-\frac{1}{2},j,k}\right)\left(\Momentum^{(x)}_{i+\frac{1}{2},j,k}+\Momentum^{(x)}_{i-\frac{1}{2},j,k}\right)\right] + p_{i,j,k},
\end{equation}
\end{widetext}
where the pressure $p$ natively resides on the cell centers. The first contribution to the advective fluxes for the momentum on faces is given by a two-point discrete divergence operator, $\mathbf{D}^{c\to f}$, involving the two nearest cell-centered fluxes, as shown in Fig.~\ref{fig0}(c). The advective flux of the normal momentum by the transverse velocities is computed at the edges of the finite-volume grid using
\begin{widetext}
\begin{eqnarray}
    \left(\uFluid^{(y)}\Momentum^{(x)}\right)_{i-\half,j-\half,k} = \frac{1}{4}\left[\left(\uFluid^{(y)}_{i,j-\half,k}+\uFluid^{(y)}_{i-1,j-\half,k}\right)\left(\Momentum^{(x)}_{i-\half,j,k}+\Momentum^{(x)}_{i-\half,j-1,k}\right)\right] \\
     \left(\uFluid^{(z)}\Momentum^{(x)}\right)_{i-\half,j,k-\half} = \frac{1}{4}\left[\left(\uFluid^{(z)}_{i,j,k-\half}+\uFluid^{(z)}_{i-1,j,k-\half}\right)\left(\Momentum^{(x)}_{i-\half,j,k}+\Momentum^{(x)}_{i-\half,j,k-1}\right)\right],
\end{eqnarray}
\end{widetext}
where the contribution of these edge-based advective fluxes to the face-based momentum is computed by a four-point discrete divergence operator $\mathbf{D}^{e\to f}$, as shown in Fig.~\ref{fig0}(c). The two operators $\mathbf{D}^{c\to f}$ and $\mathbf{D}^{e\to f}$ contribute to the total divergence of the momentum advection flux, which has been previously demonstrated to be skew-symmetric~\cite{balboa2012} for the purposes of discrete fluctuation-dissipation balance.

The computation of face-centered diffusive fluxes of heat, species and energy is straightforward by a simple interpolation of cell-centered primitive variables and transport coefficients to the faces of the grid, and the divergence of these face-centered fluxes to cell centers is computed from the standard operator $\mathbf{D}^{f\to c}$, which is the same as cell-centered numerical schemes for compressible FHD in refs.~\cite{Done1,Bala1},  as shown in Fig.~\ref{fig0}(b). A major difference between the cell-centered numerical scheme and the present staggered grid scheme arises in the discretization of the diffusive momentum flux, i.e., the viscous stress tensor $\StressTensor$. For spatially varying viscosity considered here, the diagonal terms of $\StressTensor$ corresponding to $\partial_{x}\uFluid^{(x)}$, $\partial_{y}\uFluid^{(y)}$ and $\partial_{z}\uFluid^{(z)}$ are calculated at the cell centers through a $\mathbf{G}^{f\to c}$ gradient operator for the face-centered velocities, and the off-diagonal terms corresponding to $\partial_{y}\uFluid^{(x)}$, $\partial_{z}\uFluid^{(y)}$ and $\partial_{x}\uFluid^{(z)}$ etc., are calculated at the edges through a $\mathbf{G}^{f\to e}$ gradient operator. The viscosity is appropriately interpolated for the off-diagonal terms, but requires no interpolation for the diagonal terms. Such a discretization naturally maintains the symmetric nature of the stress tensor, which is more complicated to obtain in the cell-centered approach~\cite{Bala1}. The reader is referred to Supplemental Material for details on the discrete representation of the viscous stress tensor~\cite{Note1}. Upon computing the center- and edge-based fluxes as above, the discrete divergence operators $\mathbf{D}^{c\to f}$ and $\mathbf{D}^{e\to f}$ are used to compute the divergence of the stress tensor at the faces. The viscous heating in the energy equation $\nabla\cdot\left(\StressTensor\cdot\uFluid\right)$ is calculated at the faces by appropriately averaging the center and edge-based components of $\StressTensor$ to the faces, followed by computing the divergence to the cell centers using $\mathbf{D}^{f\to c}$. 

The stochastic terms corresponding to the species flux $\widetilde{\SpeciesFlux}$ and heat flux $\widetilde{\HeatFlux}$ are calculated at the faces using the same prescription as in the cell-centered scheme for compressible FHD~\cite{Bala1}, i.e., by generating independent standard Gaussian random variables on the faces as shown in Fig.~\ref{fig0}(b), along with appropriately averaging the cell-centered transport coefficients onto the faces. However the treatment of the stochastic momentum flux, i.e., the symmetric stochastic stress tensor $\widetilde{\StressTensor}$ is different in the staggered grid scheme. Consider the symmetric matrix $\widetilde{\mathcal{Z}}$ in Eq.~\ref{stoch:Z} with six uncorrelated elements given as
\begin{equation}
    \widetilde{\mathcal{Z}} = 
    \begin{bmatrix}
    S^{xx} & S^{xy} & S^{xz} \\
    S^{xy} & S^{yy} & S^{yz} \\
    S^{xz} & S^{yz} & S^{zz}
    \end{bmatrix},
\end{equation}
where the diagonal terms are uncorrelated Gaussian random variables with variance of two, and the off-diagonal terms have a variance of one. The diagonal variables corresponding to the diagonal terms of $\widetilde{\StressTensor}$ are generated at the cell centers, whereas the off-diagonal terms are generated at the edges, as shown in Fig.~\ref{fig0}(c). Unlike the cell-centered scheme that requires a split operator approach~\cite{Done1,Bala1}, a collocated representation of the diagonal terms of $\widetilde{\StressTensor}$ in the staggered scheme makes it convenient to establish the structure of the correlation of the stochastic stress defined in Eq.~\ref{stochstress:corr}. Furthermore, because the off-diagonal stochastic stresses are generated on the same face of the staggered velocity grid, such as $\widetilde{\StressTensor}_{xz}$ and $\widetilde{\StressTensor}_{zx}$, it is easy to establish the symmetry of $\widetilde{\StressTensor}$ by generating only one random variable $S^{xz}$ corresponding to both, as shown by the edge-based location of $S^{xz}$ in Fig.~\ref{fig0}(c). The viscous heating from the stochastic stress is computed similarly to the deterministic viscous heating described above.

\subsection{\label{subsec:temporal}Temporal discretization}
Following ref.~\cite{Done1}, we use an explicit, three-stage, low-storage Runge-Kutta (RK3) scheme for integrating the FNS Eqs.~\ref{FHD}. In this scheme the stochastic terms are discretized in time in a manner that requires generating only two random fields $Z^A$ and $Z^B$ per time step, while providing a weak second-order accuracy for the noise~\cite{delong2013}. The three stages of the RK3 scheme per time step proceed as
\begin{eqnarray}
    \widetilde{\mathbf{U}}^{n+1/3} &=& \mathbf{U}^{n} + \Delta t \mathbf{R}(\mathbf{U}^{n},Z_{1}), \nonumber \\
    \widetilde{\mathbf{U}}^{n+2/3} &=& \frac{3}{4}\mathbf{U}^{n} + \frac{1}{4}\left[\widetilde{\mathbf{U}}^{n+1/3} + \Delta t \mathbf{R}(\widetilde{\mathbf{U}}^{n+1/3},Z_{2}) \right], \nonumber \\
    \mathbf{U}^{n+1} &=& \frac{1}{3}\mathbf{U}^{n} + \frac{2}{3}\left[\widetilde{\mathbf{U}}^{n+2/3} + \Delta t \mathbf{R}(\widetilde{\mathbf{U}}^{n+2/3},Z_{3}) \right], \nonumber
\end{eqnarray}
where the stochastic fluxes generated from the random fields $Z_i$ are related to each other between stages by
\begin{eqnarray}
    Z_1 &=& Z^A + \beta_1 Z^B, \nonumber \\
    Z_2 &=& Z^A + \beta_2 Z^B, \nonumber \\
    Z_3 &=& Z^A + \beta_3 Z^B.
\end{eqnarray}
We use the weights described in ref.~\cite{delong2013} where $\beta_1=\left(2\sqrt{2}+\sqrt{3}\right)/5$, $\beta_2=\left(-4\sqrt{2}+3\sqrt{3}\right)/5$, and $\beta_3=\left(\sqrt{2}-2\sqrt{3}\right)/10$.

\subsection{\label{subsec:boundary}Boundary conditions}
In addition to demonstrating the equilibrium structure of hydrodynamic fluctuations using periodic boundary conditions, in this paper we incorporate several nonperiodic boundary conditions for the fluxes of species, heat and momentum. Specifically we consider thermal (Dirichlet) and adiabatic (Neumann) boundaries for heat flux, zero concentration flux (Neumann) and fixed concentration (Dirichlet) boundaries for species flux, and no-slip (Dirichlet) and slip (Neumann) walls for tangential momentum flux. For all such non-periodic boundaries, the normal momentum vanishes at the boundary corresponding to the no net-flow condition. 

In the staggered grid for velocity, the physical boundary aligns with the faces of the grid. Therefore, unlike a cell-centered scheme for velocity, it is straightforward to set the normal velocity to zero at the walls without requiring any extra normal velocity variables to be assigned outside the domain in a ghost cell. Similarly, because the normal velocity is exactly zero at the walls, there is no requirement of prescribing any pressure value in a ghost cell, which introduces a complication in the cell-centered scheme by presenting a choice of prescribing either pressure or density in the ghost cell. Furthermore, the advective fluxes of density, concentration and energy normal to the walls are automatically set to zero at the physical boundary without requiring any special treatment.

The boundary conditions for the other cell-centered quantities, such as concentration, temperature, and tangential velocity components, that reside half a cell width away from the physical boundary are implemented using ghost cells outside the boundary. For example, a Neumann adiabatic condition at the $x=0$ boundary is implemented by setting the temperature in the ghost cell $T(-1)$ equal to the temperature in the first cell of the domain $T(0)$, which sets the heat flux from temperature gradient at $x=0$ boundary to zero. Additionally the Dufour contribution to the energy flux at $x=0$ adiabatic boundary is also set to zero.
For Dirichlet conditions such as concentration and temperature specified at the boundaries, the transport coefficients are calculated based on the boundary values, which are used to compute the fluxes at the boundaries. A modified operator using half the cell width is used to compute the divergence of these boundary fluxes to update the state of the cells in the domain adjacent to the boundaries.

\begin{figure*}[t]
\centering
\includegraphics{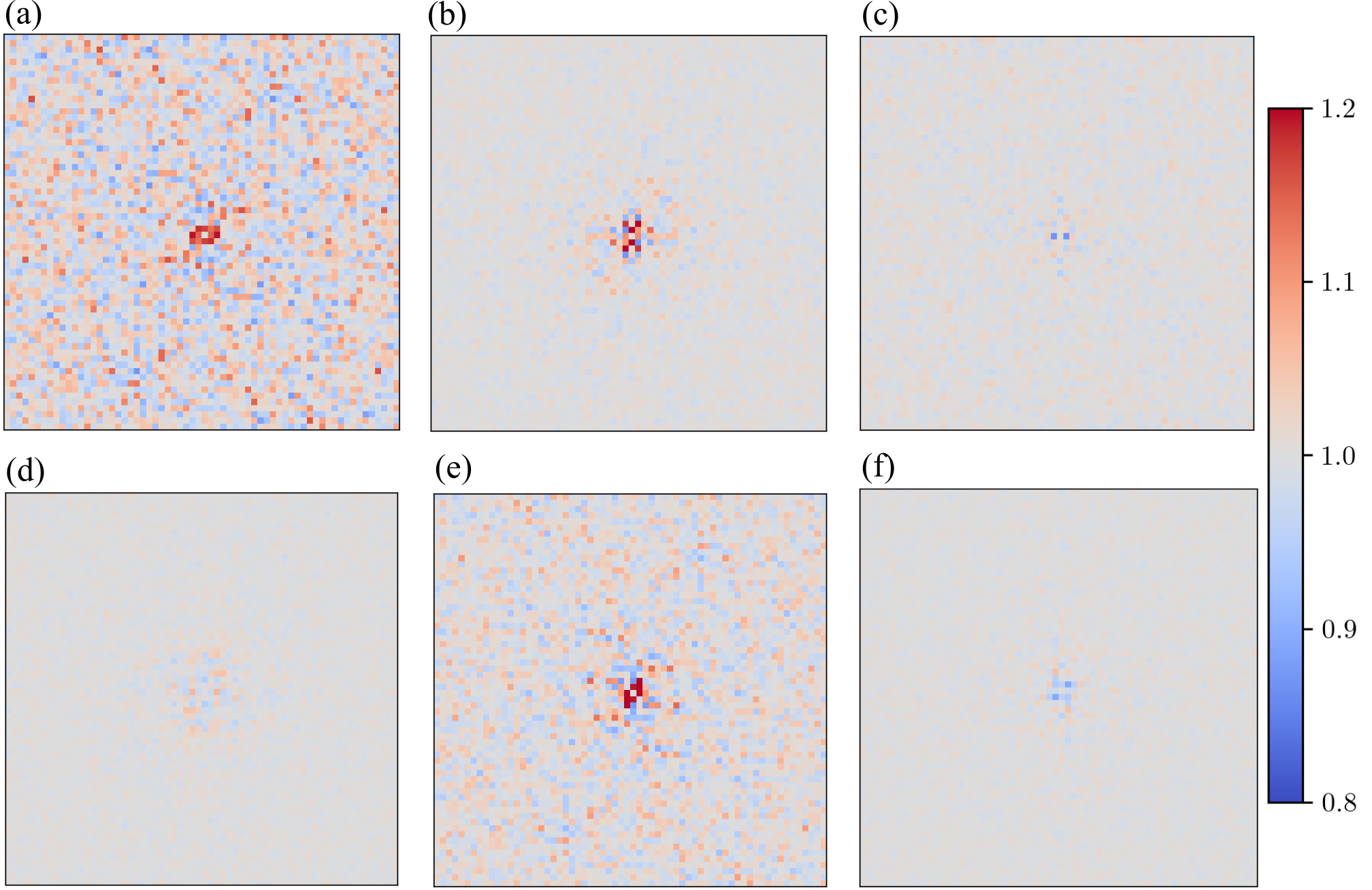}
\caption{Nondimensionalized static structure factors along the $k_y$ and $k_z$ axes for $k_x=0$ with the wavenumbers ranging from $-3.93\times10^{-7}\text{cm}^{-1}$ to $3.93\times10^{-7}\text{cm}^{-1}$. The panels correspond to (a) $\langle (\delta\widehat{\rho}) (\delta\widehat{\rho})^{*} \rangle$, (b) $\langle (\delta\widehat{\mathbf{j}^{(x)}}) (\delta\widehat{\mathbf{j}^{(x)}})^{*} \rangle$, (c) $\langle (\delta\widehat{\rho E}) (\delta\widehat{\rho E})^{*} \rangle$, (d) $\langle (\delta\widehat{\rho_{1}}) ( \delta\widehat{\rho_{1}})^{*} \rangle$, (e) $\langle (\delta\widehat{\rho_{4}}) (\delta\widehat{\rho_{4}})^{*} \rangle$, and (f) $\langle (\delta\widehat{T}) (\delta\widehat{T})^{*} \rangle$. The colored data ranges around $\pm20\%$ from the expected theoretical value of unity. The center of the image corresponds $k_y=k_z=0$.}
\label{fig1}
\end{figure*}

The treatment of stochastic fluxes at the Neumann boundaries is similar to diffusive fluxes, i.e., they are set to zero. For the stochastic fluxes at a Dirichlet boundary, the variance of the flux noise is multiplied by $\sqrt{2}$ resulting from an effective reduction of the control volume by half, as explained in detail in refs.~\cite{balboa2012,Done2}.

\section{\label{sec:results}Numerical Results}
In this section we first provide numerical evidence that the staggered scheme correctly reproduces the structure of hydrodynamic fluctuations of ideal gas mixtures at equilibrium with periodic boundary conditions, while demonstrating higher accuracy of the scheme compared to the previous cell-centered scheme~\cite{Bala1}. Next, we simulate giant fluctuations in a nonequilibrium mixture driven out of equilibrium by concentration gradients and provide comparisons of their scale-invariant, correlated structure with theoretical predictions. This is followed by various quasi-one-dimensional tests for long-ranged correlations in mixtures of real ideal gases driven by thermal and concentration gradients, particularly focusing on the role of cross-diffusion effects (such as Soret and Dufour) on these correlations that are often neglected in the theoretical analysis of such phenomena. We validate our results by comparing with DSMC simulations. Lastly, we demonstrate the ability our code to simulate the Rayleigh-Taylor instability in a two-fluid system driven by thermal fluctuations with an initially perfectly smooth interface.

\subsection{\label{sec:resultsA}Gas mixture at equilibrium: Static structure factor}
As a first test, we consider an inert mixture of four ideal noble gases with equal mass fractions at equilibrium with periodic boundaries; see Table~\ref{table1} for molecular details of these gases. The ideal gas equation of state along with an ideal gas specific heat described in Sec.~\ref{subsec:hydro} are used for the gas mixture.
\begin{table}[b]
\caption{\label{table1}
Molecular properties of species $k$ in the equilibrium gas mixture.}
\begin{ruledtabular}
\begin{tabular}{p{0.1\columnwidth}|p{0.3\columnwidth}|p{0.2\columnwidth}|p{0.2\columnwidth}|p{0.2\columnwidth}}
\textrm{$k$}&
\textrm{Species}&
\textrm{Molecular Weight (g/mol)}&
\textrm{Diameter (cm)}&
\textrm{$Y_k$}\\
\colrule
1 & Helium & $4.0026$ & $2.18\times10^{-8}$ & $0.25$ \\
2 & Neon & $20.1797$ & $2.58\times10^{-8}$ & $0.25$ \\
3 & Argon & $39.9480$ & $3.63\times10^{-8}$ & $0.25$ \\
4 & Krypton & $83.8000$ & $4.16\times10^{-8}$ & $0.25$ \\
\end{tabular}
\end{ruledtabular}
\end{table}

\begin{figure*}[t]
\centering
\includegraphics{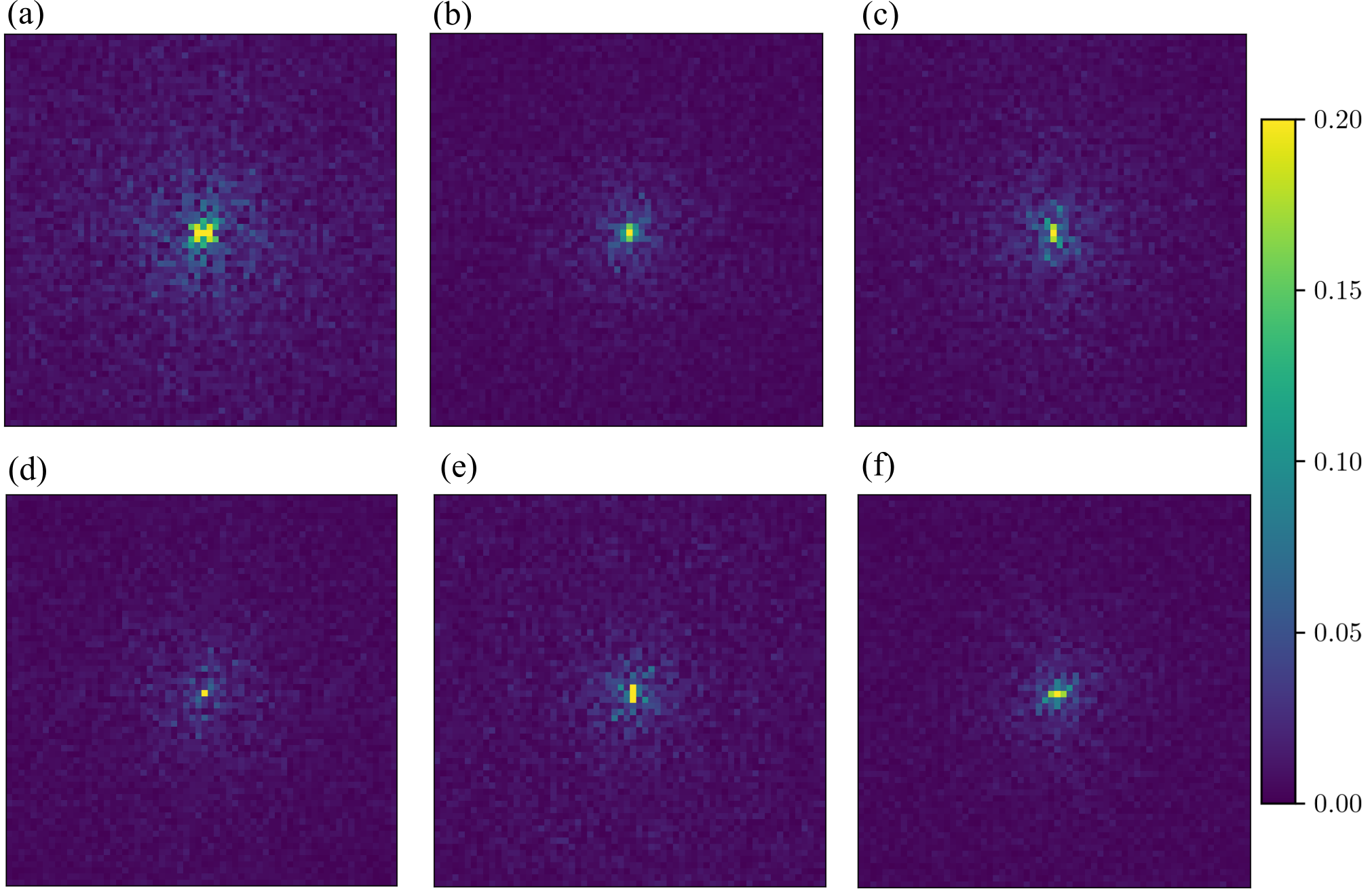}
\caption{Magnitude of nondimensionalized correlations in Fourier space (see the caption in Fig.~\ref{fig1}). The panels correspond to (a) $\langle (\delta\widehat{\rho}) (\delta\widehat{\mathbf{j}^{(x)}})^{*} \rangle$, (b) $\langle (\delta\widehat{\rho}) (\delta\widehat{T})^{*} \rangle$, (c) $\langle (\delta\widehat{\mathbf{j}^{(x)}}) (\delta\widehat{\mathbf{j}^{(y)}})^{*} \rangle$, (d) $\langle (\delta\widehat{\mathbf{j}^{(x)}}) (\delta\widehat{\rho E})^{*} \rangle$, (e) $\langle (\delta\widehat{\rho_{1}}) (\delta\widehat{\rho_{4}})^{*} \rangle$, and (f) $\langle (\delta\widehat{\uFluid^{(x)}}) (\delta\widehat{T})^{*} \rangle$.}
\label{fig2}
\end{figure*}
The simulation setup consisted of a cubic domain of size $L_x=L_y=L_z=5.12\times 10^{-4}\text{cm}$ discretized on a $64^3$ regular finite-volume grid. The system was initialized with zero velocity at ambient pressure $p=1.01\times10^{6}\text{dyn}/\text{cm}^{2}$ and temperature $T=300\text{K}$. 
The initial mass fractions of all the species are equal, i.e., $Y_i=0.25$, and the total initial density is $\rho=4.82\times10^{-4}\text{g}/\text{cm}^{3}$. A time step of $\Delta t = 10^{-12}\text{s}$ was used to advance the FHD equations, which corresponds to an acoustic Courant number $\left(||\uFluid||_{\infty}+c_{0}\right)\Delta t/\Delta x \approx 7\times 10^{-3}$, where $c_0$ is the speed of sound and $||\uFluid||_{\infty}$ is the $\ell_{\infty}$ norm of the fluid velocity vector at each point in the domain. This setup matches with the one used by the cell-centered scheme in ref.~\cite{Bala1}.

An initial simulation was run for $5\times10^{4}$ time steps to reach equilibrium after which data was collected every $10$ time steps for an additional $5\times10^{5}$ time steps. Each snapshot of the data was Fourier transformed in three dimensions, and pairwise correlations were computed in Fourier space and averaged in time to produce the static structure factor. The static structure factors were nondimensionalized by the equilibrium variance of fluctuations from classical thermodynamics~\cite{landau_statistical_1980}, as summarized in the Appendix A of ref.~\cite{Bala1}. For example, the structure factor for density fluctuations $\langle \left(\delta\widehat{\rho}\right) \left(\delta\widehat{\rho}\right)^{*} \rangle$ is normalized by $\langle \delta \rho^{2}\rangle$, where the $\delta\widehat{\rho}$ denotes the Fourier transform of the $\delta\rho$ field, the asterisk denotes the complex conjugate, and the angular brackets denote averaging over space and time. For the structure factors containing cross-correlations such as $\langle \left(\delta\widehat{\rho}\right) (\delta\widehat{T})^{*} \rangle$ that vanish at equilibrium for all wavevectors, the cross-correlations are nondimensionalized by the corresponding variances of each quantity such as $\sqrt{\langle \delta \rho^{2}\rangle \langle \delta T^{2}\rangle}$. 

Figure~\ref{fig1} shows the nondimensionalized static structure factor at equilibrium for various conserved variables as well as temperature. In a perfect numerical scheme, the nondimensionalized structure factor should be unity at all wavelengths. Our simulations show excellent agreement with theoretical predictions at all wavenumbers. Furthermore, unlike the cell-centered numerical scheme (see Fig.~1 in Balakrishnan et al.,~\cite{Bala1}), we observe significantly less statistical scatter in the structure factor for the total density $\langle (\delta\widehat{\rho}) (\delta\widehat{\rho})^{*} \rangle$ and partial density of the heaviest species $\langle (\delta\widehat{\rho_{4}}) (\delta\widehat{\rho_{4}})^{*} \rangle$ at high wavenumbers. The current scheme with the velocity on a staggered grid provides a more compact stencil for the pressure term in the momentum flux, 
which results in reduced spurious correlations of local noise that would manifest at high wavenumbers of the structure factor at finite time steps. The large statistical error observed for the lowest wave vectors results from very long relaxation times for the largest wavelengths.

Figure~\ref{fig2} shows correlations between various combinations of temperature and the conserved variables as a function of the wave number. Our results show near-zero correlations in agreement with theory that predicts that the measured quantities should be uncorrelated at equilibrium. Similar to the cell-centered numerical scheme (see Fig.~2 in~\cite{Bala1}), our staggered scheme performs well in the discretization of the stress tensor as evidenced by the excellent agreement of $\langle (\delta\widehat{\mathbf{j}^{(x)}}) (\delta\widehat{\mathbf{j}^{(y)}})^{*} \rangle$ with theory. Furthermore, our scheme outperforms the cell-centered scheme in removing spurious correlations between partial densities of lightest and heaviest species $\langle (\delta\widehat{\rho_{1}}) (\delta\widehat{\rho_{4}})^{*} \rangle$, thus demonstrating that the present numerical method robustly handles the relaxation of hydrodynamic fluctuations at most wavelengths except at the largest, which have the longest relaxation times.

\subsection{\label{sec:resultsB}Giant fluctuations and long range correlations induced by a concentration gradient}
Next we consider two examples of a binary fluid mixture at nonequilibrium induced by an external concentration gradient that exhibits giant fluctuations of the local concentration fluctuations. For the first example, we assume that the two species have identical thermodynamic and mechanical properties but can be distinguished.
The setup is similar to a previous study that utilized DSMC simulations to study diffusion enhancement due to giant fluctuations in nonequilibrium fluids~\cite{Done4}. Although this is a remarkably simplified setup compared to microgravity experiments of giant fluctuations~\cite{Vailati1}, the choice of identical but distinguishable species maintains homogeneous thermodynamic and transport properties throughout the system, where the concentration fluctuations can only diffuse within the domain. As such, this setup provides for a convenient comparison with the predictions of the linearized FHD theory~\cite{brogioli2001}.

\begin{figure}[t]
\centering
\includegraphics[width=\columnwidth]{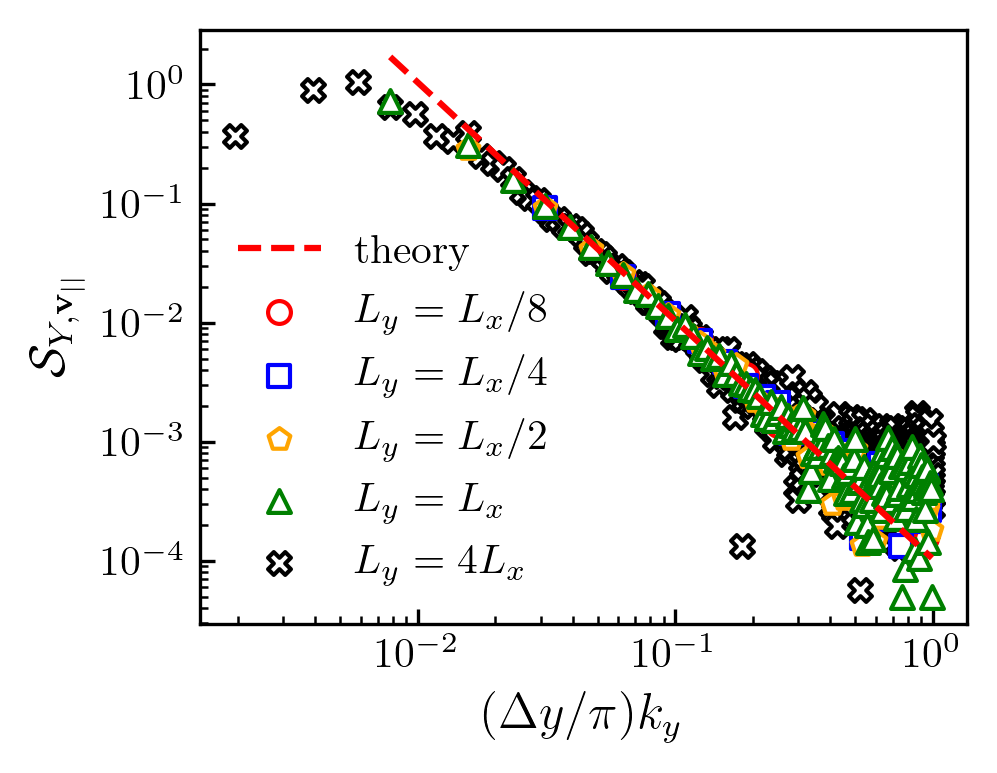}
\caption{The structure factor $\mathcal{S}_{Y, \uFluid_{||}}$ as a function of wavenumber $k_y$ normalized by cell spacing $\Delta y$ for wavevectors perpendicular to the applied concentration gradient, i.e., $k_x=0$, for various values of $L_y$. The theoretical prediction from Eq.~\ref{eq:giantfluc} for an infinitely periodic system is shown with a dotted line.}
\label{fig3}
\end{figure}

The system was initialized at $T=273\text{K}$ and $\rho=1.78\times10^{-3}\text{g}/\text{cm}^{3}$ with two identical but distinguishable species with properties corresponding to Argon (Ar). For these conditions the mean free path of the gas is $l_{\lambda}=6.26\times10^{-6}\text{cm}$. The cell sizes were $\Delta x=\Delta y=2l_{\lambda}$, and the system thickness in the $z$ direction was set as $2l_{\lambda}$ in a quasi-2D simulation (a modified formulation for the stochastic stress tensor in the quasi-2D approximation is described in Appendix~\ref{AppA}). The time step was fixed at $\Delta t=10^{-10}\text{s}$, which corresponds to an acoustic Courant number $\approx 0.2$. A concentration gradient was applied in the $x$ direction, and periodic boundary conditions were applied in the $y$ direction. The dimensions of the system were fixed at $256$ cells in $x$ direction.
We varied the number of cells in the $y$ direction from $32$ to $1024$ to study the effects of finite system size $L_y$ perpendicular to the gradient. A strong external concentration gradient was applied to the system by setting the Dirichlet values for the mass fraction of species $0$ to $Y^L=0.25$ at $x=0$ and $Y^R=0.75$ at $x=L_x$, where $L_x$ is the system length in the $x$ direction. The two boundaries in the $x$ direction were also held at a fixed temperature with no slip for the tangential velocities.

\begin{figure}[b]
\centering
\includegraphics[width=\columnwidth]{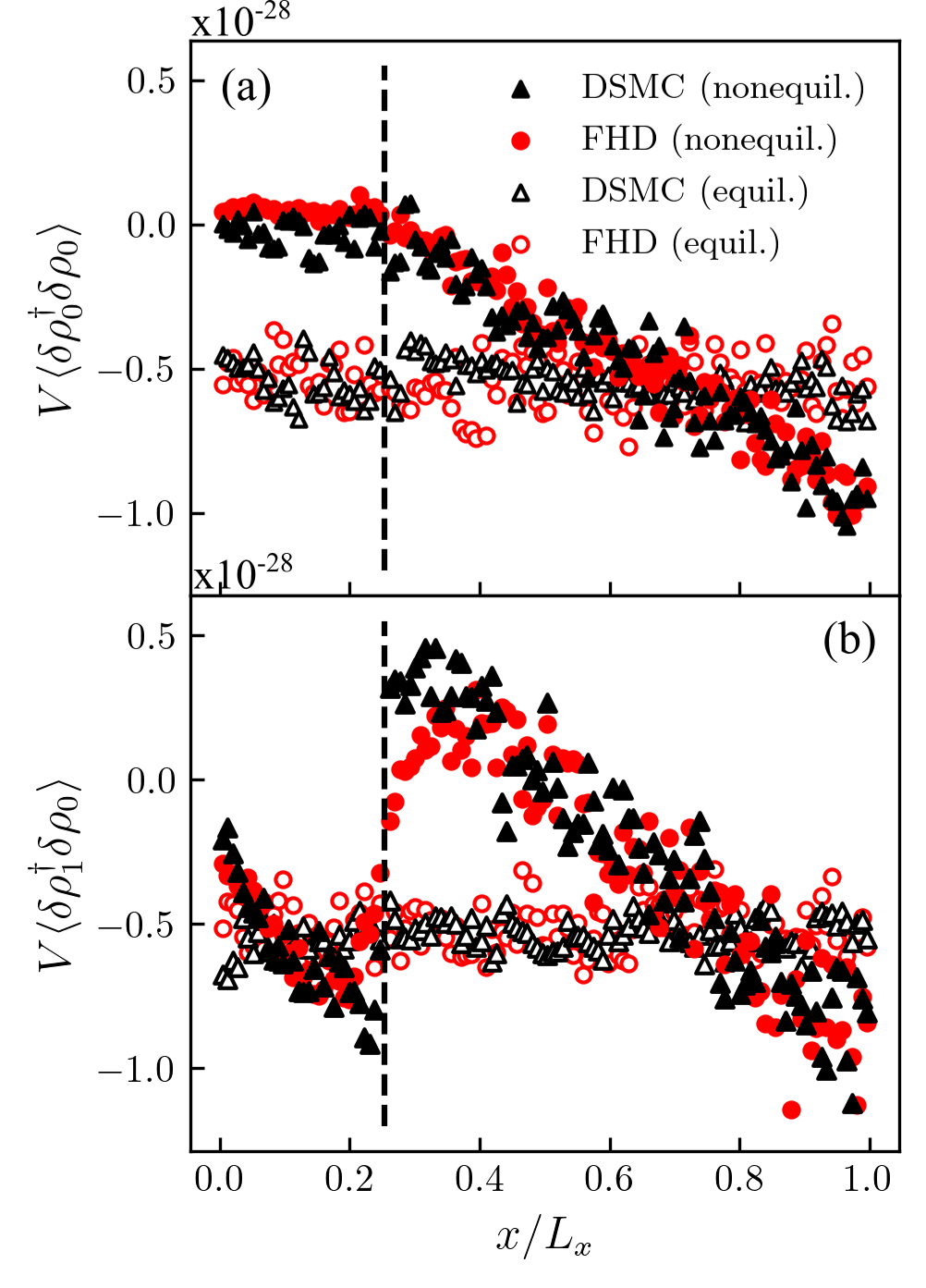}
\caption{Real-space spatial correlations in a quasi-1D binary fluid mixture. Results are shown for equilibrium and nonequilibrium steady states using FHD and DSMC simulations (see legend in (a)). (a) Spatial correlation between the fluctuations of partial density of species $0$, $\delta\rho_{0}^{\dagger}$,  at $x^{\dagger}$ with the fluctuations $\delta\rho_{0}$ in every other cell. (b) Spatial correlation between the fluctuations of partial density of species $1$, $\delta\rho_{1}^{\dagger}$, at $x^{\dagger}$ with the fluctuations $\delta\rho_{0}$ in every other cell. The correlations are normalized by the volume $V$ of either FHD or DSMC cell in their respective cases. The position of $x^{\dagger}$ is shown by vertical dashed lines, and the value of the correlations at $x^{\dagger}$ have been removed in (a) and (b).}
\label{fig4}
\end{figure}

After an initial run for $10^{6}$ time steps to relax the system to a statistically stationary state, snapshots of the data were collected at every $10$ time steps for at least an additional $5\times10^{6}$ time steps to compute the Fourier-space spatial correlations between concentration fluctuations $\delta Y$ and velocity fluctuations $\delta\uFluid_{||}$ in the $x$ direction parallel to the concentration gradient. The data was further averaged from an ensemble of two simulations that were run in parallel for each case of varying $L_y$. The linearized equations of FHD predict long-ranged correlations, $\mathcal{S}_{Y, \uFluid_{||}} = \langle (\delta\widehat{Y})(\delta\widehat{\uFluid_{||}})^{*}\rangle$, between $\delta Y$ and $\delta\uFluid_{||}$ resulting in the giant fluctuation phenomena, and whose magnitude scales linearly with the magnitude of the applied gradient $|\nabla\overline{ Y}|$~\cite{brogioli2001}
\begin{equation}
    \mathcal{S}_{Y, \uFluid_{||}} = -\frac{k_B T}{(\eta+\rho D)|\bm{k}|^{2}} \left(\text{sin}^{2}\theta\right) |\nabla\overline{ Y}|,
    \label{eq:giantfluc}
\end{equation}
where $|\bm{k}|$ is the magnitude of the wavevector $\bm{k}$, $\text{sin}^{2}\theta = k_{\perp}^2/k^2$, i.e., $\theta$ is the angle between $\bm{k}$ and the vector of the applied concentration gradient $\nabla\overline{ Y}$, and $D$ is the binary diffusion coefficient.

Figure~\ref{fig3} shows the structure factor $\mathcal{S}_{Y, \uFluid_{||}}$ for various values of $L_y$ along the periodic direction. Excellent agreement is observed between theory and our numerical results. The correlations exhibit a $k_{y}^{-2}$ power-law decay at low wavenumbers, which is indicative of their long-ranged and scale-invariant nature.
Because the system consists of a mixture of identical gases, these long-ranged correlations result entirely from the system being out of thermodynamic equilibrium since there is no net macroscopic transport. The giant fluctuations have previously been demonstrated to enhance the diffusive transport in nonequilibrium fluids through the advection of concentration by thermal velocity fluctuations~\cite{Done3,Done4}. At very low wavenumbers, we observe a suppressed correlation compared to theory, resulting from the effect of confining walls in the $x$ direction that occurs at wavenumbers $k_y$ comparable to $2\pi/L_x$~\cite{Done4}. The numerical structure factor $\mathcal{S}_{Y, \uFluid_{||}}$ in Fig.~\ref{fig3} is in excellent agreement with previous DSMC simulations of a similar nonequilibrium system~\cite{Done4}.

In the second example we analyze the nonequilibrium effect of a strong applied concentration gradient on the long-ranged correlations of hydrodynamic fluctuations in real space. For this example, we consider a quasi-1D simulation of a binary mixture of identical but labeled Neon (Ne) gas (see Appendix~\ref{AppA} for a modified formulation of the stochastic stress tensor in the quasi-1D approximation). The system domain is a $128\times1\times1$ grid with cubic cells of size $3\times10^{-6}\text{cm}$. The gas mixture was initialized at a temperature of $T=300\text{K}$ and density $\rho=8.17\times10^{-4}\text{gm}/\text{cm}^{3}$ with an equimolar concentration of both the identical species. The two walls confining the gas mixture prescribe a Dirichlet condition for concentration and temperature.

A strong concentration gradient was imposed on the system by setting the concentration of species $0$ at the boundaries at low and high $x$ coordinates to $Y^{L}=0.1$ and $Y^{R}=0.9$ respectively, while the wall temperature was fixed at $300\text{K}$.  A time step of $\Delta t=10^{-12}\text{s}$ was used to march the solution forward in time, which corresponds to an acoustic Courant number $\approx 1.5 \times 10^{-2}$. The simulation was initially run for $2\times10^{7}$ time steps to reach a statistically stationary state, after which it was run for an additional $1.3\times10^{8}$ time steps to gather statistics. \commentout{\MarginPar{Discussion seems a bit unclear.  Previous paragraph talks about concentration gradient but first simulation is equilibrium- JBB. \Srivastava{I mention both eq \& noneq now}}}An ensemble of $16$ simulations were run in parallel to provide sufficiently large statistics for computing correlations between various fluctuating hydrodynamic quantities. We also modeled this system using DSMC since it accurately describes nonequilibrium transport for strong gradients, such as in shock waves. The reader is referred to Appendix~\ref{AppB} for details about the DSMC simulations. Equilibrium simulations using the same setup but without an applied concentration gradient were also conducted both in DSMC and continuum FHD for the purposes of comparison with the nonequilibrium case.

Figure~\ref{fig4}(a) shows real-space correlation between the fluctuations of partial density of species $0$, i.e., $\delta\rho_{0}^{\dagger}=\delta(\rho Y_{0})^{\dagger}$, at $x^{\dagger}=L_x/4-\Delta x/2$ with the fluctuations $\delta\rho_{0}$ everywhere else in the domain. At equilibrium, the real-space correlation of these fluctuations normalized by the cell volume, i.e., $V\langle\delta\rho_{0}^{\dagger} \delta\rho_{0}\rangle$, is theoretically predicted to be $1.36\times10^{-26}\text{gm}^{2}/\text{cm}^{3}$ at $x=x^{\dagger}$, and $-5.3\times10^{-29}\text{gm}^{2}/\text{cm}^{3}$ at every other $x\neq x^{\dagger}$ due to effects of mass conservation. We observe excellent correspondence between the theoretical predictions and the results from both DSMC and continuum FHD simulations in Fig.~~\ref{fig4}(a). As expected, beyond the variance introduced by mass conservation, there is no spatial correlation between the fluctuations of $\rho_0$ at equilibrium. At equilibrium, a similar correlation $\langle\delta\rho_{1}^{\dagger}\delta\rho_{0}\rangle$ between the fluctuations of partial density of species $1$, $\delta\rho_{1}^{\dagger}$, at $x^{\dagger}$ and $\delta\rho_{0}$ at every other $x$ is observed, as shown in Fig.~\ref{fig4}(b), which corresponds well with theoretical predictions.

Upon the application of a concentration gradient, long-ranged correlations in both $\langle\delta\rho_{0}^{\dagger} \delta\rho_{0}\rangle$ and $\langle\delta\rho_{1}^{\dagger} \delta\rho_{0}\rangle$ emerge at the nonequilibrium steady state, as shown in Fig.~\ref{fig4}. Remarkably, our staggered grid scheme for solving FHD equations agrees excellently with DSMC simulations even with strong concentration gradients that are typically well-resolved by DSMC but often pose a challenge for continuum methods.

\subsection{\label{sec:resultsC}Nonequilibrium fluctuations induced by temperature gradient: Role of Soret and Dufour effects}
In the next example, we consider a binary mixture of dissimilar fluids under a strong applied thermal gradient induced by constant temperature walls. Because the two gases are physically dissimilar, thermal gradients in the system will introduce concentration gradients (Soret effect), and these concentration gradients will induce a heat flux (Dufour effect) in addition
to the heat flux induced by temperature gradients~\cite{de1962non}. Several previous experiments using light scattering~\cite{croccolo2012,bataller2017} and shadowgraphy techniques~\cite{croccolo2019,vailati2020,garcia-fernandez2022} have studied the effect of induced concentration gradients due to an applied temperature gradient (Soret effect) on the nonequilibrium fluctuations in binary and ternary fluid mixtures.

\begin{figure}[t]
\centering
\includegraphics[width=\columnwidth]{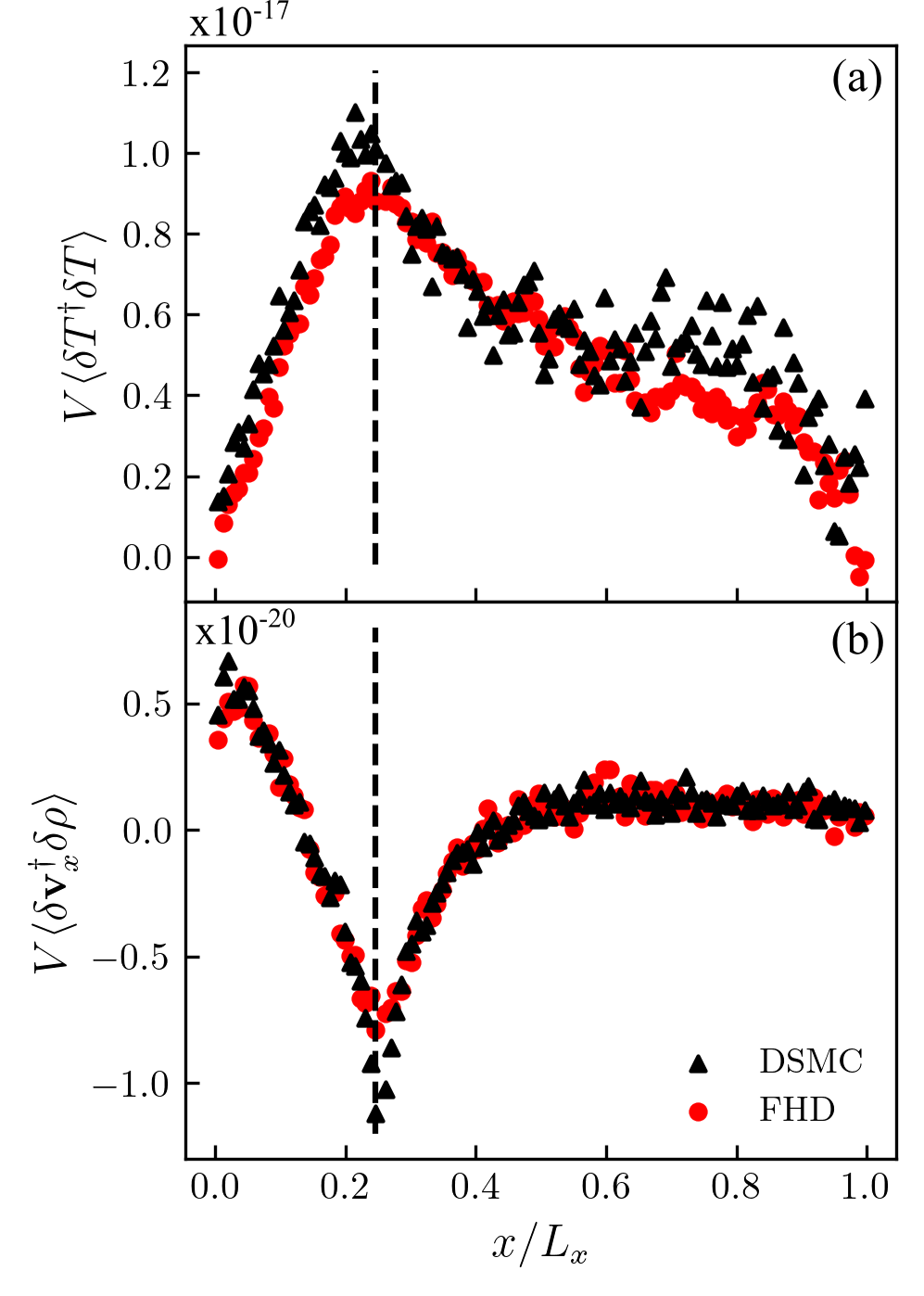}
\caption{Real-space spatial correlations in a quasi-1D binary mixture of dissimilar gases under thermal gradient. Results are shown for FHD and DSMC simulations (see legend in (b)). (a) Spatial correlation between the fluctuations of temperature $\delta T^{\dagger}$ at $x^{\dagger}$ with the fluctuations $\delta T$ in every other cell. (b) Spatial correlation between the fluctuations of $x$-velocity $\delta \uFluid_{x}^{\dagger}$ at $x^{\dagger}$ with the fluctuations of total density $\delta\rho$ in every other cell. The correlations are normalized by the volume $V$ of either FHD or DSMC cell in their respective cases. The position of $x^{\dagger}$ is shown by vertical dashed lines, and the value of the correlations at $x^{\dagger}$ have been removed in (a) and (b).}
\label{fig5}
\end{figure}

Previous theoretical analyses using linearized FHD equations for the nonequilibrium fluctuations in fluid mixtures induced by the Soret effect have neglected the role of Dufour effect, and assumed incompressibility of flow and constant thermodynamic properties that are independent of temperature and the composition of the mixture (Boussinesq approximation)~\cite{ortizdezarate2014,ortizdezarate2020}. Although these assumptions are well-justified for liquids, they break down in the case of gas mixtures, which is our focus here. Particularly, we are interested in demonstrating the role of the Dufour effect on nonequilibrium fluctuations in ideal gas mixtures. Furthermore, we do not make any assumptions about the spatially-varying thermodynamic properties of the mixture.

\begin{figure}[t]
\centering
\includegraphics[width=\columnwidth]{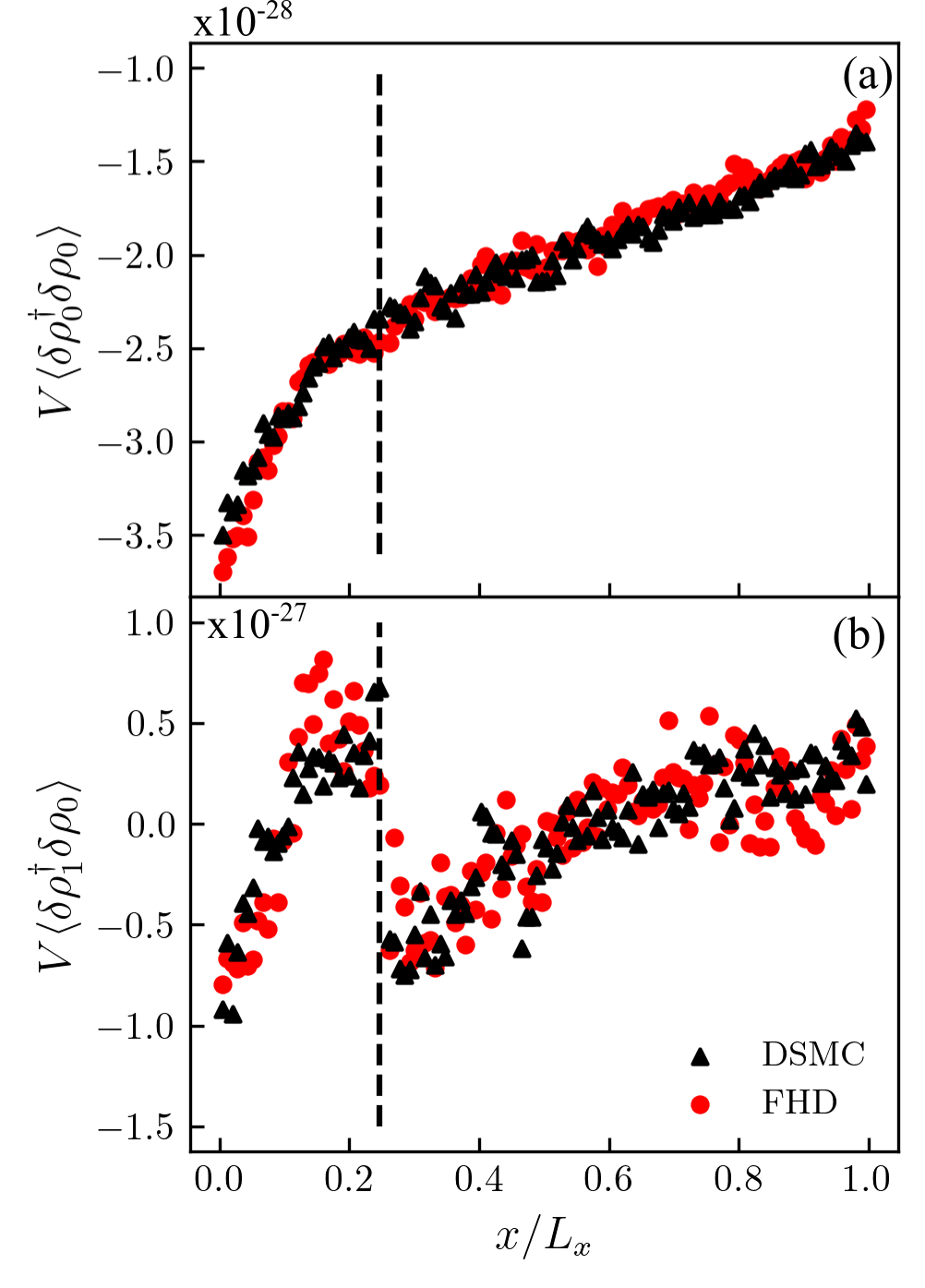}
\caption{Real-space spatial correlations in a quasi-1D binary mixture of dissimilar gases under thermal gradient. Results are shown for FHD and DSMC simulations (see legend in (b)). (a) Spatial correlation between the fluctuations of partial density of species $0$, $\delta \rho_{0}^{\dagger}$, at $x^{\dagger}$ with the fluctuations $\delta \rho_{0}$ in every other cell. (b) Spatial correlation between the fluctuations of partial density of species $1$, $\delta \rho_{1}^{\dagger}$, at $x^{\dagger}$ with the fluctuations $\delta \rho_{0}$ in every other cell. The correlations are normalized by the volume $V$ of either FHD or DSMC cell in their respective cases. The position of $x^{\dagger}$ is shown by vertical dashed lines, and the value of the correlations at $x^{\dagger}$ have been removed in (a) and (b).}
\label{fig6}
\end{figure}

We consider a binary mixture of two dissimilar inert gases, Neon (Ne; species $0$) and Krypton (Kr; species $1$), confined between two thermal walls in a quasi-1D simulation. The length of the domain $L_x=3.84\times10^{-4}\text{cm}$ was discretized over a $128\times1\times1$ grid with cubic cells. The gas mixture was initialized at a temperature of $T=273\text{K}$ and density $\rho=1.45\times10^{-3}\text{gm}/\text{cm}^{3}$ with an equal mass concentration of both the species. The two walls confining the gas mixture prescribed a Dirichlet condition for temperature, and prescribed no flux of normal momentum and concentration. A time step of $\Delta t=10^{-12}\text{s}$ was used to advance the solution, which corresponds to an acoustic Courant number $\approx 1.5 \times 10^{-2}$. At $t=0$, the wall at $x=L_x$ was set at a temperature at $519\text{K}$, while the wall at $x=0$ remained at $273\text{K}$, resulting in a strong thermal gradient of $\sim 6\times10^{5}\text{Kcm}^{-1}$. The simulation was initially run for $2\times10^{7}$ time steps to achieve a nonequilibrium steady state, after which it was run for an additional $1.3\times10^{8}$ time steps to gather statistics. An ensemble of $4$ simulations were run in parallel to provide sufficiently large statistics of the correlations between various fluctuating hydrodynamic quantities. We also ran DSMC simulations for this system to verify and corroborate the results from FHD numerical solution. The reader is referred to Appendix~\ref{AppB} for details about the DSMC simulations.

Figure~\ref{fig5}(a) shows the long-ranged correlation between temperature fluctuations at nonequilibrium under a thermal gradient, which is a classic signature in mesoscale fluids driven out of equilibrium by temperature gradients~\cite{Mans1}. Fig.~\ref{fig5}(b) shows the long-ranged correlations between velocity and density fluctuations, and excellent agreement is observed between DSMC and numerical solutions of FHD using the staggered grid. Furthermore, previous studies using a cell-centered grid scheme demonstrated that continuum FHD significantly under-predicts the sharp peak in the correlation $\langle\delta \uFluid_{x}^{\dagger}\delta\rho\rangle$ near $x^{\dagger}$ to almost half of the value obtained from DSMC simulations~\cite{Bell1,LadigesMembrane19}. The staggered numerical scheme provides a better correspondence with DSMC as a result of the compact discretization of stochastic momentum fluxes, although it still under-predicts the correlation near $x^{\dagger}$ to almost three-quarters of the value obtained from DSMC simulations. We note that because the velocities and momenta reside on the faces of the finite-volume grid in the staggered grid scheme, the fluctuations in the velocity $\delta \uFluid_{x}^{\dagger}$ are computed by averaging the fluctuations in the momentum $\delta \mathbf{j}_{x}$ from the neighboring faces to the center of the cell~\cite{Note1}, which is a numerical artifact that could also contribute to the under-prediction.

\begin{figure}[b]
\centering
\includegraphics[width=\columnwidth]{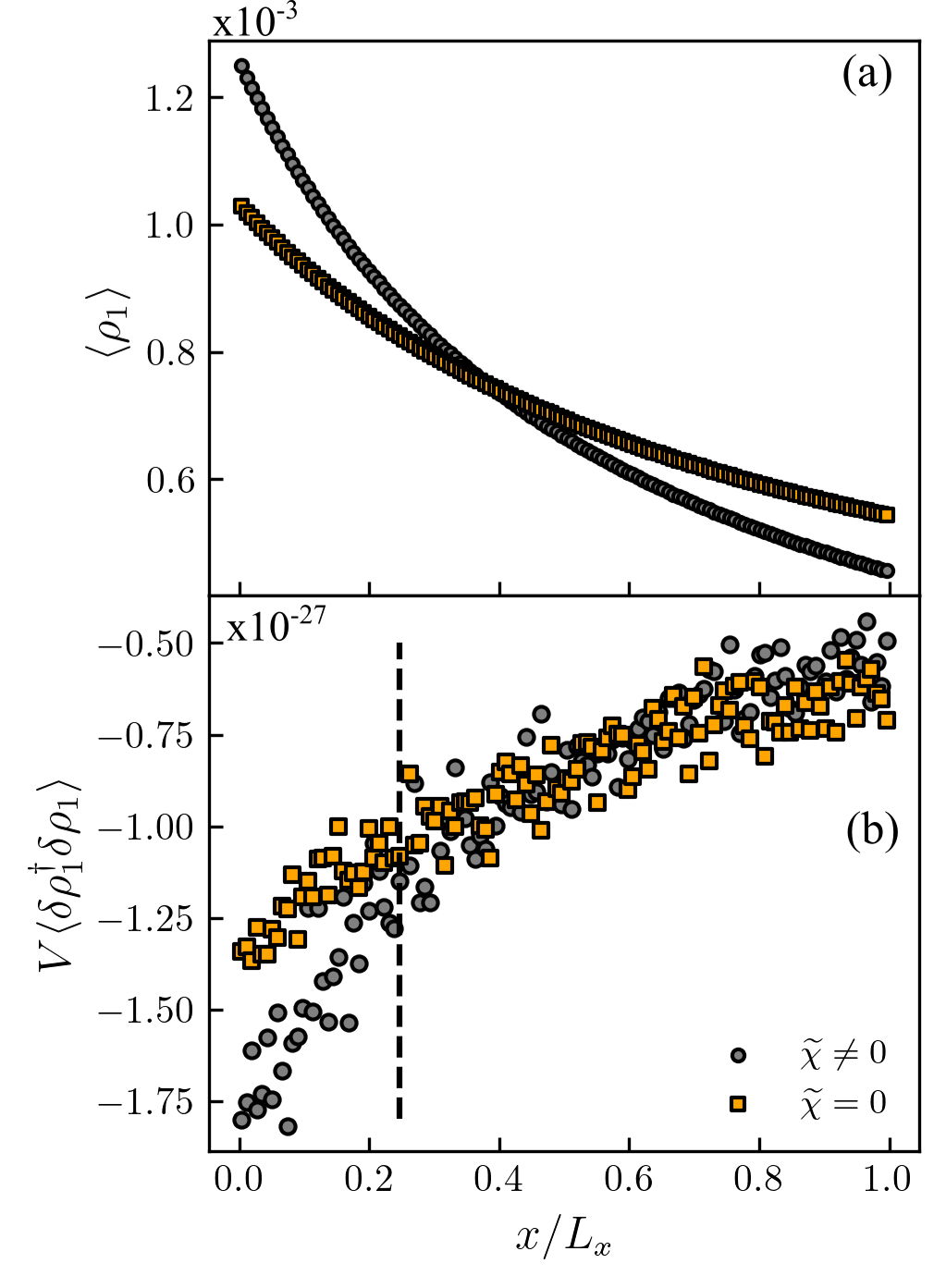}
\caption{The role of Soret and Dufour effects in the mean and fluctuating hydrodynamics of a mixture of dissimilar gases under an applied thermal gradient. Results are shown for FHD simulations with $\widetilde{\chi} \neq 0$ (circles), and $\widetilde{\chi} = 0$ (squares) that corresponds to neglecting the Soret and Dufour effects. (a) The mean profile $\langle \rho_{1} \rangle$ of the partial density of species $1$. (b) Spatial correlation between the fluctuations of partial density of species $1$, $\delta \rho_{1}^{\dagger}$, at $x^{\dagger}$ with the fluctuations $\delta \rho_{1}$ in every other cell. The correlations are normalized by the volume $V$ of the FHD cell. The position of $x^{\dagger}$ is shown by vertical dashed lines, and the delta function for correlation at $x^{\dagger}$ has been removed in (b).}
\label{fig7}
\end{figure}

\begin{figure*}[t]
\centering
\includegraphics{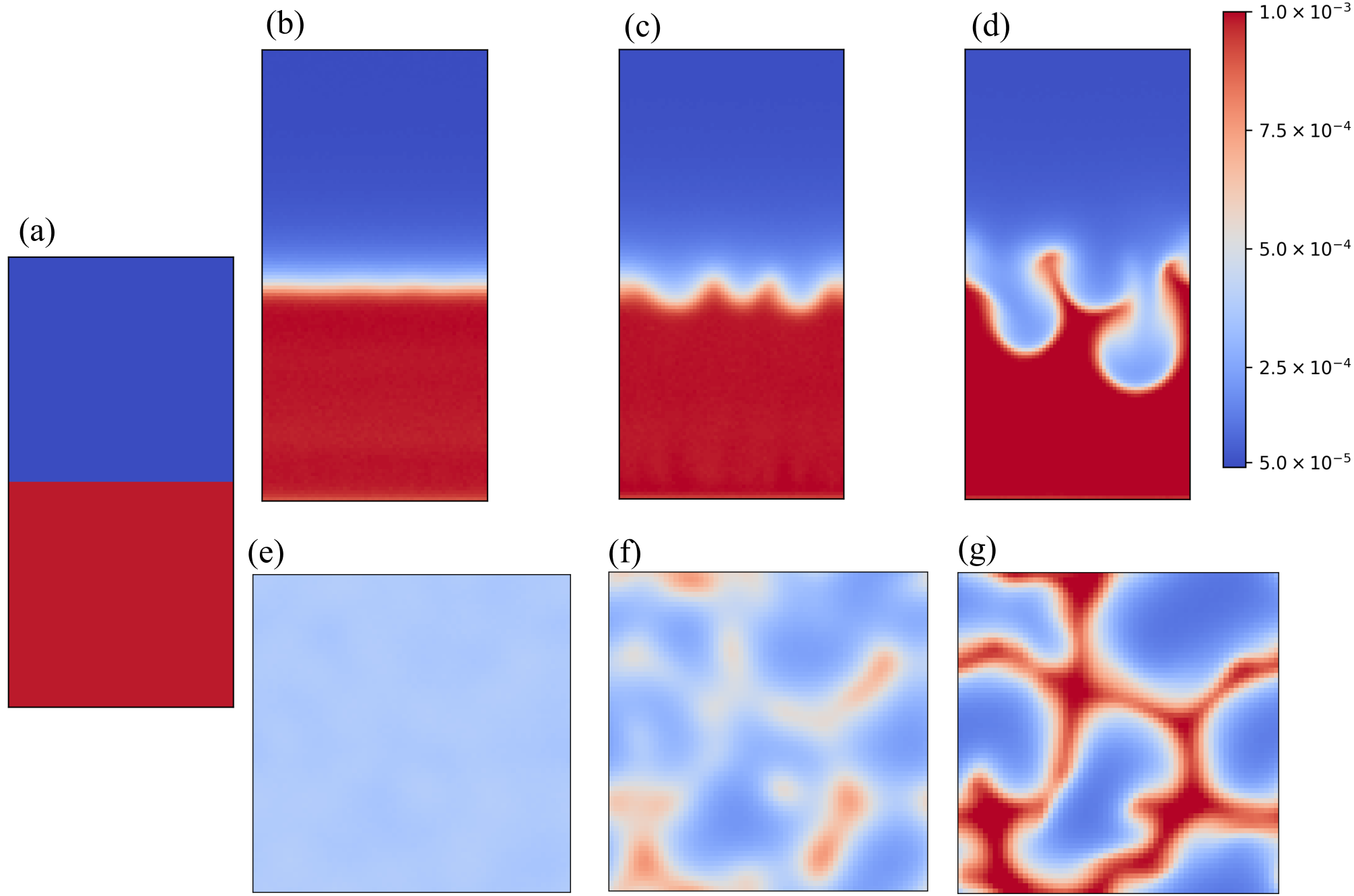}
\caption{Rayleigh-Taylor instability. The images depict the temporal evolution of the partial density of the lighter species $\rho_{0}$. Panels (a) - (d) show the vertical cross section of $\rho_0$ at times $t=0$, $t=2.0\times10^{-8}\text{s}$, $t=3.75\times10^{-8}\text{s}$ and $t=5.0\times10^{-8}\text{s}$ respectively. Panels (e) - (f) show the horizontal slice at the center of the domain corresponding to (b) - (d). The data range for all the images is shown in the color bar on the top right of the figure.}
\label{fig8}
\end{figure*}

Next we consider the fluctuations of the partial densities of each species at nonequilibrium under the action of external thermal gradient. The induced concentration gradient (or the species \emph{separation}) introduces long-ranged correlation between concentration fluctuations, as evidenced by $\langle\delta \rho_{0}^{\dagger} \delta \rho_{0}\rangle$ and $\langle\delta \rho_{1}^{\dagger} \delta \rho_{0}\rangle$ in Figs.~\ref{fig6}(a) and (b) respectively. The $\langle\delta \rho_{0}^{\dagger} \delta \rho_{0}\rangle$ correlation has a constant (in space) contribution from the species conservation with an additional nonequilibrium component arising from the induced gradients in concentrations. Excellent agreement is observed between FHD and DSMC simulations. The correlation for $\langle\delta \rho_{1}^{\dagger} \delta \rho_{0}\rangle$ is noisier than $\langle\delta \rho_{0}^{\dagger} \delta \rho_{0}\rangle$ resulting from the relatively lower mole fraction (or number density) of the heavier species.

We ascertain the role of Dufour contribution to the energy flux and Soret contribution to the species flux on the long-ranged hydrodynamic correlations by simulating the same system as above and setting $\widetilde{\chi} = 0$ in Eqs.~\ref{eq:specflux_ideal}, \ref{heat_ideal}, and \ref{eq:stochfluxes}. As shown in Fig.~\ref{fig7}(a), the mean profile of $\langle \rho_{1} \rangle$ differs significantly when $\widetilde{\chi} = 0$ as there is no additional Soret contribution to the species flux from the temperature gradient. Similarly, the mean profile of $\langle T \rangle$ also differs upon neglecting the Dufour contribution to the heat flux, and the reader is referred to the Supplemental Material for details~\cite{Note1}. Concomitantly, the structure of the hydrodynamic fluctuations is also modified upon the absence of Soret and Dufour effects. Fig.~\ref{fig7}(b) shows the long-ranged correlation between the fluctuations in $\rho_{1}$, given by $\langle\delta \rho_{1}^{\dagger}\delta\rho_{1}\rangle$, for the two cases of $\widetilde{\chi} = 0$ and $\widetilde{\chi} \neq 0$. The magnitude of $\langle\delta \rho_{1}^{\dagger}\delta\rho_{1}\rangle$ is weaker when $\widetilde{\chi} = 0$, and this results from a weaker local gradient in $\rho_{1}$, since the magnitude of these correlations scales as $\left(\nabla \rho_{1}\right)^{2}$~\cite{ortizdezarate2014}. Similar differences are observed in the correlation of temperature fluctuations $\langle \delta T^{\dagger} \delta T \rangle$, and the reader is referred to the Supplemental Material for the details~\cite{Note1}. Therefore, the Soret and Dufour effects that emerge from the Onsager's reciprocity arguments~\cite{de1962non} are important to be considered while analyzing both mean and fluctuating hydrodynamics of fluid mixtures at nonequilibrium, particularly in the case of compressible gas mixtures as described here.

\subsection{\label{sec:resultsD}Hydrodynamic instability induced by thermal fluctuations: Rayleigh-Taylor instability}
As a final example, we demonstrate that thermal fluctuations can trigger a Rayleigh-Taylor instability in mesoscale fluid mixtures with an perfectly smooth initial interface separating a heavy fluid above a light fluid. Even in the absence of any macroscopic perturbations, the molecular-scale thermal fluctuations, which are modeled in a coarse-grained sense by FHD, trigger the development and growth of the instability. Because such thermal fluctuations contain a whole spectrum of wavelengths of perturbation, the perturbations corresponding to the unstable wavelength~\cite{sharp1984} grow the instability quickly, whereas the perturbations corresponding to other wavelengths diffuse at the interface. This has been reported in particle simulations and here we confirm those observations by FHD~\cite{kadau2004,gallis2016}.

We model a mixture of two fictitious monoatomic gases consisting of a lighter species of molar mass $m_0=10\text{g}/\text{mol}$ and a heavier species of molar mass $m_1=100\text{g}/\text{mol}$. The two species have the same molecular diameter $d_1=d_2=9.63\times10^{-8}\text{cm}$. The simulation domain was defined on a $64\times64\times128$ grid with $\Delta x=\Delta y = \Delta z = 1.6\times10^{-5}\text{cm}$. The low and high $z$ boundaries prescribed Neumann boundary conditions for concentration, heat flux (adiabatic walls) and tangential momentum (full slip walls), and periodic boundaries were assigned in the other two directions. Two layers of the gas mixture were initialized such that upper half contains a heavier composition of the mixture $Y_0=0.3$ and $Y_1=0.7$, and the lower half contains a lighter composition of the mixture $Y_0=0.7$ and $Y_1=0.3$, corresponding to an Atwood number of $0.82$~\cite{sharp1984}. The interface separating the two mixture halves was initially perfectly flat as shown by the variation of the partial density of the lighter species $\rho_0$ in Fig.~\ref{fig8}(a). The density and temperature at the top $z$ wall were set at $\rho=1.4\times10^{-3}\text{g}/\text{cm}^{3}$ and $T=300\text{K}$ respectively. The pressure, density and temperature were varied throughout the $z$ direction such that the system was in hydrostatic equilibrium with the gravity, which was set to $g=10^{13}\text{cm}/\text{s}^{2}$. As is commonly done in molecular simulations, an exceptionally large value of $g$ was chosen to speed up the initiation of fluid instability. The transport and thermodynamic properties of the gas mixture were computed based on the prescription by Giovangigli~\cite{Giov1}.

We fixed the time step set of the simulation at $\Delta t = 2.5\times10^{-13}\text{s}$, which corresponds to an acoustic Courant number $\approx 10^{-2}$. The simulation initially proceeded with a stratified diffusion of the species across the interface resulting from the concentration gradients, as shown by the variation of $\rho_0$ at an early time in Figs.~\ref{fig8}(b) and (e). Soon after, an instability started growing at the interface with a characteristic wave-like pattern, as seen in Figs.~\ref{fig8}(c) and (f). At later times, the instability developed into growing bubble- and spike-like penetrating structures~\cite{gallis2016}, as seen in Figs.~\ref{fig8}(d) and (g). The Mach number of the fluid flow during these later times of instability growth was $\sim 0.1$. Remarkably, this instability was triggered in a simulation starting from a perfectly flat interface, and it results entirely from thermal fluctuations. By contrast, in conventional Navier-Stokes simulations an artificial perturbation of the interface is needed to initiate the hydrodynamic instability.

\section{\label{sec:conclusions}Conclusions and Future Work}
A staggered grid numerical scheme for the solution of compressible, multispecies, non-isothermal, fluctuating Navier-Stokes equations was demonstrated to more accurately reproduce both equilibrium and nonequilibrium hydrodynamic fluctuations of fluids mixtures when compared to our previous cell-centered scheme~\cite{Bala1}. By simulating multispecies mixtures of ideal gases, we demonstrated that cross-diffusion effects, such as Soret and Dufour, importantly govern the long-ranged correlations of hydrodynamic fluctuations, and these observations were validated against DSMC simulations. The present numerical scheme was demonstrated to be adept in simulating thermal fluctuation-driven fluid instabilities such as Rayleigh-Taylor. As a part of our future work, we will utilize the accuracy of our numerical scheme to study the role of thermal fluctuations in compressible turbulent flows by simulating the fluctuating hydrodynamics of Taylor-Green vortex flows.\cite{Gallis2022, bell_nonaka_garcia_eyink_2022}

The numerical scheme accurately represents various boundary conditions such as isothermal/adiabatic walls, slip/no-slip walls, and walls with fixed values of species concentrations. In future we will formulate numerical methods to model particle reservoirs at the boundaries within the staggered numerical scheme to simulate an open system. A previous DSMC study improvised standard reservoir models to avoid non-physical correlations in hydrodynamic fluctuations~\cite{tysanner2005}, and similar care will be required while formulating reservoir boundaries in the present compressible FHD numerical scheme. These developments will also contribute to our ongoing and future work on coupling FHD with surface chemistry models to simulate chemical reactions at a reactive surface, such as in catalysis. Similarly, we also plan to include bulk chemistry models in the present numerical scheme to model reacting, multispecies mixtures towards simulating chemical processes such as combustion and explosive detonation, where thermal fluctuations have been shown to be important~\cite{Lemarchand2}. An isothermal, incompressible solver for the FHD equations of reactive fluid mixtures was developed previously, and will be extended to include non-isothermal and compressibility effects using the staggered grid scheme described here~\cite{kim2018fluctuating}.

In earlier work we used kinetic theory to formulate stochastic models to simulate the transpiration of a gas across a nanoporous membrane at the nanoscale~\cite{LadigesMembrane19}, with practical applications in gas separation and sensing technologies~\cite{wang2017c}. A single-species, cell-centered numerical scheme for compressible, non-isothermal FHD was used for this purpose, and good agreement was observed between DSMC and continuum solutions. As a part of our ongoing work, we are reformulating the kinetic theory models for transpiration in the staggered grid numerical scheme described here to more accurately investigate the long-ranged correlations of hydrodynamic fluctuations across a nanoporous membrane, particularly in the presence of multiple species, which is of practical importance.

Building on this formulation, we also plan to develop a hybrid DSMC-continuum algorithm to provide higher fidelity modeling of the fluctuating hydrodynamics near the nanoporous membrane. Such an algorithm specifically involves implementing a DSMC representation of the fluid mixture near the membrane, which is coupled to a continuum, compressible FHD representation of the bulk fluid domain away from the membrane. The hybrid coupling strategy will be based on an approach developed previously in~\cite{garcia1999adaptive}, but will need to be generalized to include multiple species using the compressible, multispecies, staggered-grid scheme described here. The resulting hybrid algorithm will provide a more accurate representation of the fluid mixture within the Knudsen layer near the membrane, which was found to be inaccurately represented in a continuum FHD description~\cite{LadigesMembrane19}.

\begin{acknowledgments}
This work was supported by the U.S. Department of Energy, Office of Science, Office of Advanced Scientific Computing Research, Applied Mathematics Program under contract No. DE-AC02-05CH11231. This research used resources of the National Energy Research Scientific Computing Center, a DOE Office of Science User Facility supported by the Office of Science of the U.S. Department of Energy under Contract No. DE-AC02-05CH11231.
\end{acknowledgments}

\appendix

\section{\label{AppA}Quasi-1D and quasi-2D FHD simulations}
In quasi-1D and quasi-2D simulations, the fluxes along the neutral direction are zeroed out. For e.g., in a quasi-1D simulation with transport along the $x$ direction, all energy, species and momentum fluxes along $y$ and $z$ direction are set to zero. Although this is straightforward to do for species and energy fluxes, care is required to set the diagonal terms of the stochastic stress tensor that produce the correct covariances for the stochastic stress.

Consider the components of the viscous stress tensor $\StressTensor$ for the specific case of zero bulk viscosity (valid for monoatomic gases~\cite{Giov1})
\begin{equation}
\Pi_{ij} = -\eta \left( \frac{\partial u_i}{\partial x_j} + \frac{\partial u_j}{\partial x_i}  \right) + \delta_{ij} \left( \frac{2}{3} \eta {\bf \nabla} \cdot {\bf u} \right).
\label{B1}
\end{equation} In three dimensions, the components of the stochastic stress tensor $\widetilde{\StressTensor}$ is~\cite{Espa1}
\begin{equation}
    \widetilde{\StressTensor} = \sqrt{2k_B T \eta} \widetilde{\mathcal{Z}} -
\frac{\sqrt{2k_B \eta T}}{3} \mathrm{Tr} ( \widetilde{\mathcal{Z}} ) \mathbb{I},
\label{B2}
\end{equation}
where the diagonal terms of the Gaussian tensor $\widetilde{\mathcal{Z}}$ have a variance of two and the off-diagonal terms have a variance of unity. The covariances of the diagonal components of the stochastic stress tensor are
\begin{widetext}
\begin{eqnarray}
    \langle \widetilde{\Pi}_{xx}(\mathbf{r},t) \widetilde{\Pi}_{xx}(\mathbf{r}',t') \rangle &=& \left(\frac{8\eta k_{B} T}{3} \right)_{\mathbf{r},t} \delta(\mathbf{r}-\mathbf{r}')\delta(t-t'), \nonumber \\
    \langle \widetilde{\Pi}_{xx}(\mathbf{r},t) \widetilde{\Pi}_{yy}(\mathbf{r}',t') \rangle &=& -\left(\frac{4\eta k_{B} T}{3} \right)_{\mathbf{r},t}  \delta(\mathbf{r}-\mathbf{r}')\delta(t-t').
\label{B3}
\end{eqnarray}
\end{widetext}

In quasi-1D case, all the terms of viscous stress $\StressTensor$ and stochastic stress $\widetilde{\StressTensor}$ are zero except $\Pi_{xx}$ and $\widetilde{\Pi}_{xx}$ respectively. The viscous stress is given as:
\begin{equation}
    \Pi_{xx} = -2 \eta u_{x} + \frac{2}{3} \eta u_x,
    \label{B4}
\end{equation}
where ${\bf \nabla} \cdot {\bf u} = u_x$ and $u_x = \partial u/\partial x$. However, $\widetilde{\Pi}_{xx}$ is now given as: 
\begin{equation}
    \widetilde{\Pi}_{xx} = \sqrt{3k_B T \eta} \widetilde{\mathcal{Z}}_{xx} -
\sqrt{\frac{k_B \eta T}{3}} \widetilde{\mathcal{Z}}_{xx},
\label{B5}
\end{equation}
to give the same covariance $\langle \widetilde{\Pi}_{xx}(\mathbf{r},t) \widetilde{\Pi}_{xx}(\mathbf{r}',t') \rangle$ in Eq.~\ref{B3}. Here, we have used $\mathrm{Tr} ( \widetilde{\mathcal{Z}} ) = \widetilde{\mathcal{Z}}_{xx}$. Note that $\widetilde{\Pi}_{xx}$ for quasi-1D differs from the three-dimensional case given in Eq.~\ref{B2} by a factor of $\sqrt{3/2}$.

For the case of quasi-2D, the stochastic stress tensor is determined by the following. We write it as:
\begin{equation}
    \widetilde{\StressTensor} = a\sqrt{2k_B T \eta} \widetilde{\mathcal{Z}} -
b\frac{\sqrt{2k_B \eta T}}{3} \mathrm{Tr} ( \widetilde{\mathcal{Z}} ) \mathbb{I},
\label{B6}
\end{equation}
where the numerical weights $a$ and $b$ are to be determined. In quasi-2D, $\widetilde{\Pi}_{xx}$ and $\widetilde{\Pi}_{yy}$ are:
\begin{eqnarray}
    \widetilde{\Pi}_{xx} = a\sqrt{2k_B T \eta} \widetilde{\mathcal{Z}}_{xx} -
b\frac{\sqrt{2k_B \eta T}}{3} ( \widetilde{\mathcal{Z}}_{xx} + \widetilde{\mathcal{Z}}_{yy} ) , \nonumber \\
    \widetilde{\Pi}_{yy} = a\sqrt{2k_B T \eta} \widetilde{\mathcal{Z}}_{yy} -
b\frac{\sqrt{2k_B \eta T}}{3} ( \widetilde{\mathcal{Z}}_{xx} + \widetilde{\mathcal{Z}}_{yy} ) ,
\label{B7}
\end{eqnarray}
where $a$ and $b$ are real numbers that satisfy the covariances in Eq.~\ref{B3}. The following values satisfy this
\begin{eqnarray*}
    a &=& 1, \\
    b &=& \frac{3+\sqrt{3}}{2},
\end{eqnarray*}
for Gaussian-distributed $\widetilde{\mathcal{Z}}_{xx}$ and $\widetilde{\mathcal{Z}}_{yy}$ with a variance of 2.0. Note that because $a=1$, the off-diagonal terms of $\widetilde{\StressTensor}$ remain unchanged from their three-dimensional version.

The modified formulation of the stochastic stress tensor described above for quasi-1D and quasi-2D simulations requires generating and storing fewer random numbers than a full 3D simulation, which makes numerical scheme more computationally efficient.

\section{\label{AppB}DSMC simulations}
The results from the numerical solutions to the FHD equations are compared with molecular simulations performed using the direct simulation Monte Carlo (DSMC) method, which has traditionally been used to simulate gas dynamics at both macroscopic and molecular scales. The reader is referred to Ref.~\cite{Bird1}, and to refs.~\cite{Alex2,Garc2} for a pedagogical treatment of this method. The simulation begins with an initial random placement of DSMC particles (each representing a single gas atom or a collection of atoms) in the domain with velocities drawn from a Maxwell-Boltzmann distribution. In each time step, the particles are first advected along their velocity without accounting for any collisions between them. At this stage, any appropriate boundary conditions are also imposed. Subsequently, the collisions between particles are chosen by a stochastic process that conserves momentum and energy, with the post-collision velocities selected from kinetic theory distributions. The equilibrium and non-equilibrium fluctuations in DSMC accurately reproduce the physical spectra of spontaneous thermal fluctuations; this has been confirmed previously by excellent agreement with fluctuating hydrodynamic theory~\cite{Garcia1987,Mans1} and molecular dynamics simulations~\cite{mareschal1992}.

In the implementation used here we have employed the no time counter (NTC) method \cite{Bird1} to select collisions, with the particles treated as hard spheres, and each representing a single gas atom. This produces the same ideal gas transport properties used for the FHD simulations, discussed in Sec.~\ref{subsec:species}. Particle reflections from boundaries are either diffuse, corresponding to an isothermal no-slip boundary, or specular, corresponding to an adiabatic full slip boundary; see Sec.~\ref{subsec:boundary}. The DSMC method has been implemented using the AMReX framework~\cite{zhang2019a}.

For the DSMC simulations described in Sec.~\ref{sec:resultsB} the same parameters were used as FHD simulations, with the following exceptions: (a) DSMC simulations used non-cubic cells, where $\Delta x =  3 \times 10^{-6}$cm (same as FHD), but $\Delta y = \Delta z = 3.5 \times 10^{-6}$cm, and (b) a time step of $\Delta t = 3 \times 10^{-11}$s was used in the DSMC simulations. One simulator particle was used to represent each real gas molecule, resulting in approximately 115,000 particles; this ensures a minimum of 100 particles of each species per cell. To impose the concentration boundary conditions, particles impacting the wall have their species randomly assigned proportionally to the specified concentration. Due to Knudsen effects~\cite{Bird1,kundt1875ueber} that are significant far from equilibrium, it is generally not possible to impose a strict Dirichlet condition on the hydrodynamic and thermodynamic fluid properties at a boundary in DSMC simulations. This occurs because the bulk fluid properties are obtained by integrating over all particle velocities, while the boundary condition is applied only to particles with velocities moving away from the wall. This typically results in a `slip' at the boundary, where the bulk fluid properties undershoot the condition being applied to the outgoing particles. To compensate for this effect, the probability of a particle being assigned to species $0$ at the left and right boundaries was set as $P^{L}=0.06$ and $P^{R}=0.94$, which yielded boundary concentrations similar to the FHD values of $Y^{L}=0.1$ and $Y^{R}=0.9$, as described in Sec.~\ref{sec:resultsB}.

For the DSMC simulations described in Sec.~\ref{sec:resultsC}, the same cell size and time step was used as the DSMC simulations in Sec.~\ref{sec:resultsB}. This yields approximately 100,000 particles representing neon, and 25,000 particle representing krypton, resulting in a minimum of 100 particles of each species per cell. The thermal gradient is imposed through Dirichlet temperature walls, and a temperature slip occurs at the boundaries, similar to the case of the applied concentration gradient. The wall temperatures of $534$K and $261$K were set in the DSMC simulations, which resulted in bulk fluid temperatures that matched well with the $519$K and $273$K temperature boundary conditions used in the FHD simulations, as described in Sec.~\ref{sec:resultsC}.

\bibliography{references}

\ifarXiv
    

\section*{Supplementary material}

## I. ERRATA FOR BALAKRISHNAN ET AL., PHYS. REV. E, 89:013017, 2014

We have identified the following errors in Balakrishnan *et al.*, Phys. Rev. E, 89:013017, 2014. The corresponding corrections are:

- Eq.7: $\sqrt{\frac{2k_B \eta T}{3}}$ should change to $\frac{\sqrt{2k_B \eta T}}{3}$

- Eq. 29 (part 2): $\mathbf{\Pi}$ should be instead $\nabla \cdot \mathbf{\Pi}$ on the left hand side

- Eq. 30: $D^{f-c}$ should instead be $G^{c-f}$, and $G^{c-f}$ should be $D^{f-c}$

- Eq. 31: $D^{n-c}$ should instead be $D^{c-n}$, and $G^{c-n}$ should be $G^{n-c}$

- Eq. 32: RHS should be

$$G^{n-c}(\eta D^{c-n} v) + D^{n-c}(\eta (G^{c-n} v)^T)$$

- The noise term for heat flux $\widetilde{Q}_x$ should have a $T^2$ in the square root after $k_B$

- Eq. 27: The hyperbolic energy flux should be: $\mathbf{v}(\rho E + P)$

## II. COMPARISON OF VARIOUS ADVECTION SCHEMES

The advective fluxes of species density and energy on the faces can alternatively be defined using momentum on the staggered grid and interpolating the thermodynamic variables from the cell centers to the faces. For example, the advective flux of species density $\mathbf{F}_H^{\mathbf{Y}}$ can be defined as

$$(\mathbf{F}_H)_{i+1/2,j,k}^{\mathbf{Y},(x)} = \frac{1}{2} [\mathbf{Y}_{i+1,j,k} + \mathbf{Y}_{i,j,k}] \mathbf{j}_{i+\frac{1}{2},j,k}^{(x)}, \tag{1}$$

and the advective flux of energy $\mathbf{F}_H^E$ can be defined in the following two ways:

$$(\mathbf{F}_H)_{i+1/2,j,k}^{E,(x)} = \frac{1}{2} \left[ (E + p/\rho)_{i+1,j,k} + (E + p/\rho)_{i,j,k} \right] \mathbf{j}_{i+\frac{1}{2},j,k}^{(x)}, \quad \text{or} \tag{2}$$

$$(\mathbf{F}_H)_{i+1/2,j,k}^{E,(x)} = \frac{1}{2} \left[ \left( \sum_l Y_l h_l + \mathbf{v} \cdot \mathbf{v}/2 \right)_{i+1,j,k} + \left( \sum_l Y_l h_l + \mathbf{v} \cdot \mathbf{v}/2 \right)_{i,j,k} \right] \mathbf{j}_{i+\frac{1}{2},j,k}^{(x)}, \tag{3}$$

where the thermodynamic variables $p$, $\rho$, $T$ and $Y_l$ are interpolated from the cell centers to the faces.

We compare the three advection schemes in terms of hydrodynamic fluctuations in gas mixtures in equilibrium and nonequilibrium settings. The first scheme (method A) corresponds to Eq. 27 of the main text for discretizing the advective fluxes of energy and species density. The second scheme (method B) corresponds to Eq. 1 of the Supplemental Material for discretizing the advective fluxes of species density, and Eq. 2 of the Supplemental Material for discretizing the advective fluxes of energy density. The third scheme (method C) corresponds to Eq. 1 of the Supplemental Material for discretizing the advective fluxes of species density, and Eq. 3 of the Supplemental Material for discretizing the advective fluxes of energy density.

The equilibrium structure factors for various hydrodynamic quantities, as described in Section IV A, are computed using method B and method C, and shown in Figs. S1 and S2 respectively. Similarly, the nondimensionalized correlation between various hydrodynamic quantities are shown in Figs. S3 and S4 respectively. Evidently, the choice of advection scheme does not impact the equilibrium structure factors and hydrodynamic correlations.

In Fig. S5 we show various real-space hydrodynamic correlations for the three advection schemes in a binary gas mixture under an applied thermal gradient, as described in the Section IV C of the main text.

## III. DISCRETE REPRESENTATION OF THE VISCOUS STRESS TENSOR

The components of the stress tensor $\mathbf{\Pi}$ are expressed in the matrix form as:

$$\mathbf{\Pi} = -\eta \begin{bmatrix} 2\mathbf{v}_x^{(x)} & \mathbf{v}_y^{(x)} + \mathbf{v}_x^{(y)} & \mathbf{v}_z^{(x)} + \mathbf{v}_x^{(z)} \\ \mathbf{v}_y^{(x)} + \mathbf{v}_x^{(y)} & 2\mathbf{v}_y^{(y)} & \mathbf{v}_z^{(y)} + \mathbf{v}_y^{(z)} \\ \mathbf{v}_z^{(x)} + \mathbf{v}_x^{(z)} & \mathbf{v}_z^{(y)} + \mathbf{v}_y^{(z)} & 2\mathbf{v}_z^{(z)} \end{bmatrix} - (\kappa - \frac{2}{3}\eta)(\nabla \cdot \mathbf{v})\mathbb{I}$$

The diagonal terms $\mathbf{v}_x^{(x)}$, $\mathbf{v}_y^{(y)}$, $\mathbf{v}_z^{(z)}$, and $\nabla \cdot \mathbf{v}$ are computed at the cell centers as:

$$\mathbf{v}_{x,i,j,k}^{(x)} = \frac{\mathbf{v}_{i+1/2,j,k}^{(x)} - \mathbf{v}_{i-1/2,j,k}^{(x)}}{\Delta x}, \tag{4}$$

with analogs for $\mathbf{v}_y^{(y)}$ and $\mathbf{v}_z^{(z)}$, where the face-centered velocities $\mathbf{v}$ are computed using the staggered momentum density at the faces and material density in the neighboring cell centers, and described in the main text. The shear and bulk viscosities reside at the cell centers. The off-diagonal components of $\mathbf{\Pi}$ are computed on the centers of the edges of the

3

finite volume cell as:

$$(\mathbf{v}_y^{(x)} + \mathbf{v}_x^{(y)})_{i+1/2,j+1/2,k} = \frac{\mathbf{v}_{i+1/2,j+1,k}^{(x)} - \mathbf{v}_{i+1/2,j,k}^{(x)}}{\Delta y} + \frac{\mathbf{v}_{i+1,j+1/2,k}^{(y)} - \mathbf{v}_{i,j+1/2,k}^{(y)}}{\Delta x} \quad (5)$$

with analogs for other terms in the off-diagonal terms. The viscosity at the center of the edge of the finite volume cell is computed as:

$$\eta_{i+1/2,j+1/2,k} = \frac{\eta_{i+1,j+1,k} + \eta_{i+1,j,k} + \eta_{i,j+1,k} + \eta_{i,j,k}}{4} \quad (6)$$

The divergence $\nabla \cdot \mathbf{\Pi}$ for the viscous momentum flux is computed by the divergence operator $\mathbf{D}^{c \to f}$ for the diagonal components of $\mathbf{\Pi}$, and the operator $\mathbf{D}^{e \to f}$ for the off-diagonal components.

## IV. MEAN PROFILES AND HYDRODYNAMIC FLUCTUATIONS IN BINARY GAS MIXTURE UNDER THERMAL GRADIENT

To further evaluate the role of Soret and Dufour effect, we computed the mean profiles and real-space correlations of various hydrodynamic quantities in a binary gas mixture under an applied thermal gradient, as described in Sec. IV C of the main text. Figs. S6(a) and (c) show the profile of mean density of species 0 and temperature for the two cases: with Soret and Dufour effects $\widetilde{\chi} \neq 0$, and without these effects $\widetilde{\chi} = 0$. Minor variations are observed between the two cases. Similar minor variations are seen in the real-space correlations of temperature fluctuations $\delta T$, and species 0 density fluctuations $\delta \rho_0$, as shown in Figs. S6(b) and (d).

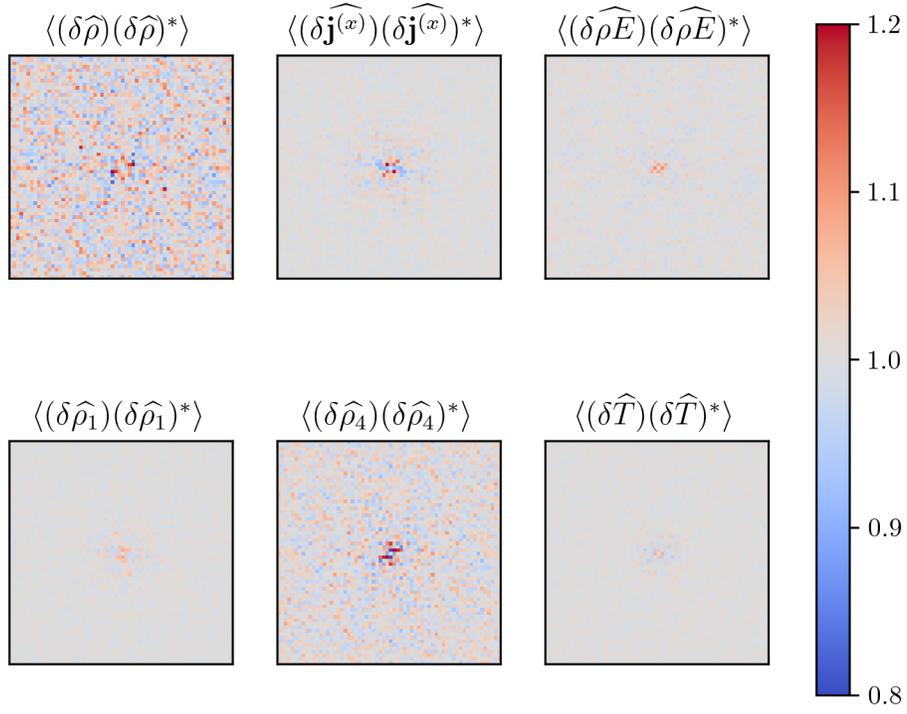

FIG. S1. Nondimensionalized static structure factors for method B at equilibrium. Refer to Fig. 2 of the main text for the description of each panel.

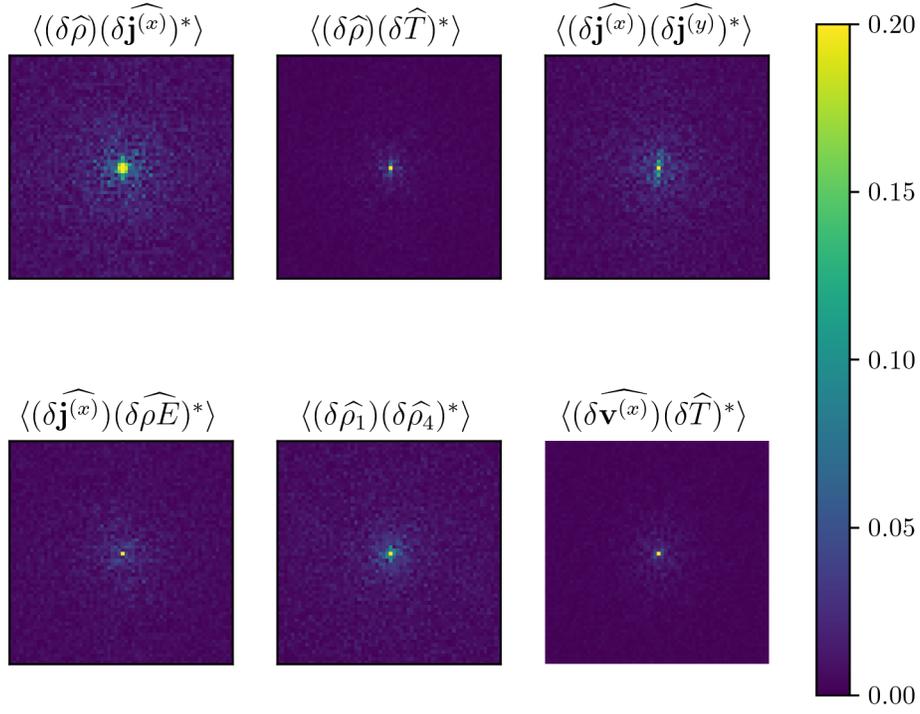

FIG. S2. Nondimensionalized static structure factors for method C at equilibrium. Refer to Fig. 2 of the main text for the description of each panel.

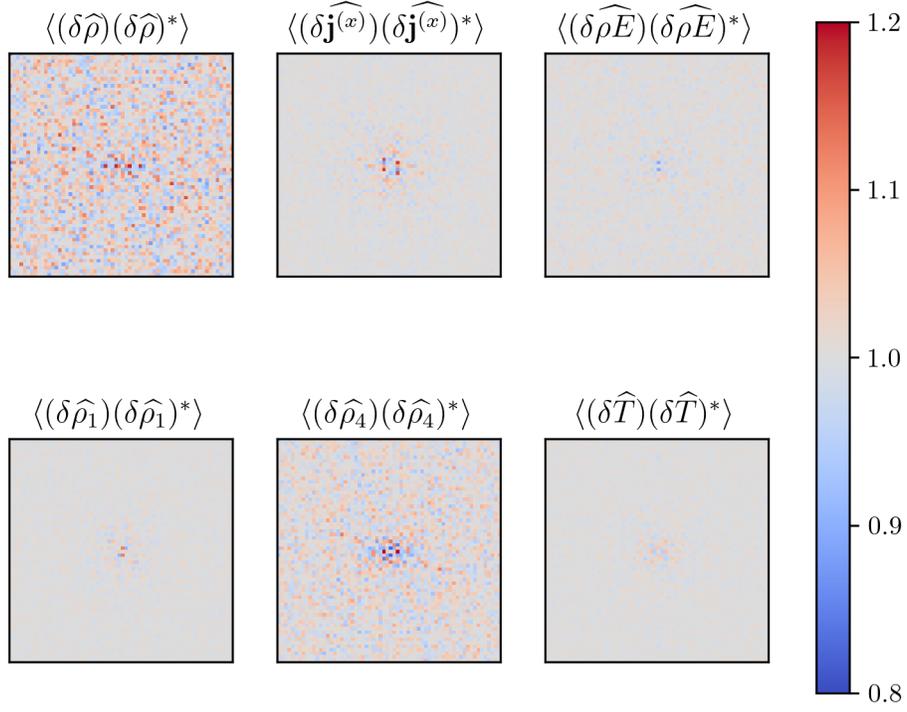

FIG. S3. Magnitude of nondimensionalized correlations in Fourier space at equilibrium for method B. Refer to Fig. 2 of the main text for the description of each panel.

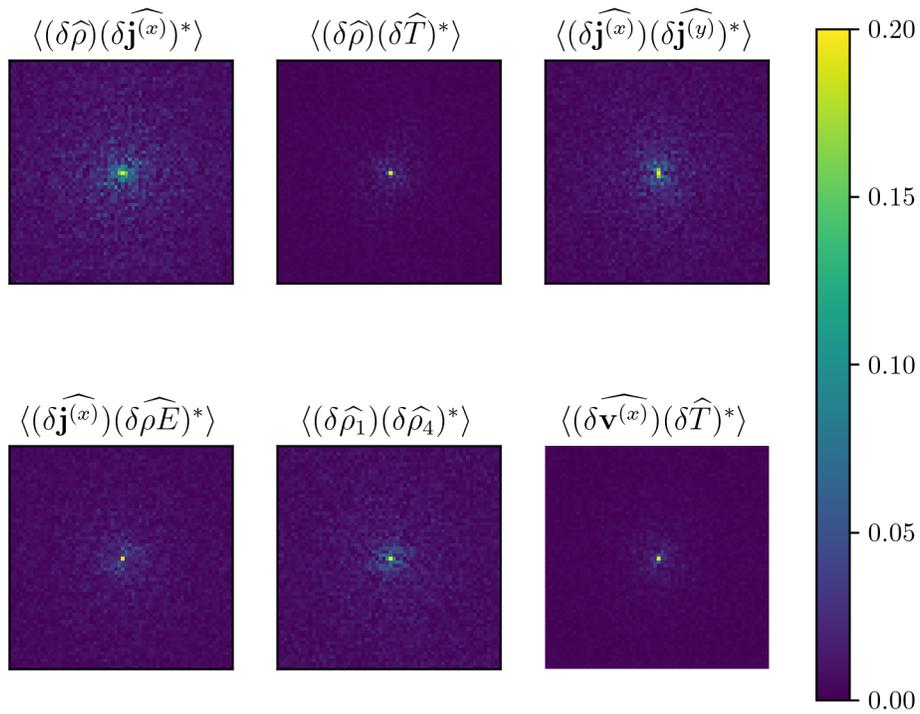

FIG. S4. Magnitude of nondimensionalized correlations in Fourier space at equilibrium for method C. Refer to Fig. 2 of the main text for the description of each panel.

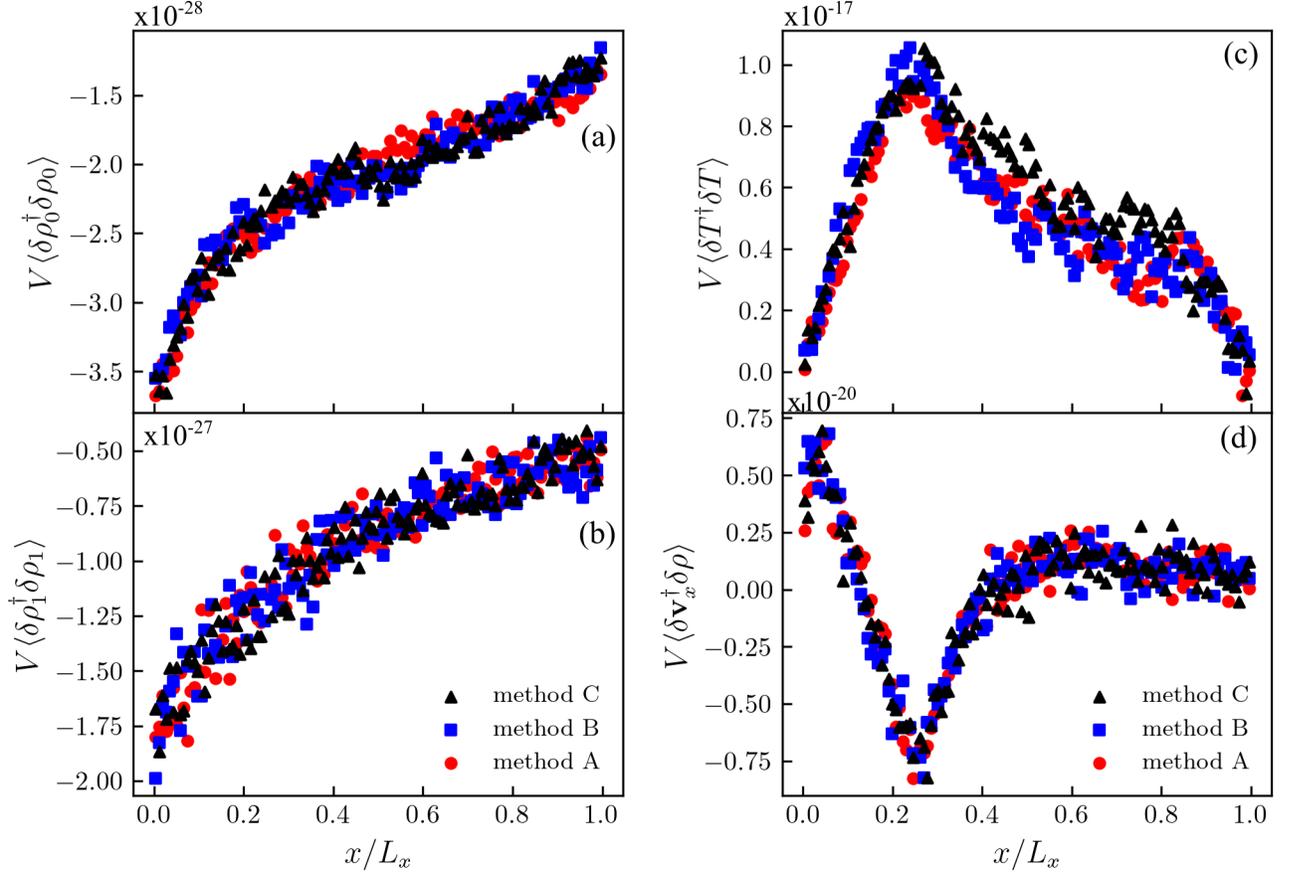

FIG. S5. Real-space spatial correlations in a quasi-1D binary mixture of dissimilar gases under thermal gradient for the three advection schemes (see legends in (b) and (d)). (a) $\langle \delta\rho_0^\dagger \delta\rho_0 \rangle$, (b) $\langle \delta\rho_1^\dagger \delta\rho_1 \rangle$, (c) $\langle \delta T^\dagger \delta T \rangle$, and (d) $\langle \delta\mathbf{v}_x^\dagger \delta\rho \rangle$. The correlations are normalized by the volume $V$ of the FHD cell. The value of the correlations at $x^\dagger$ have been removed.

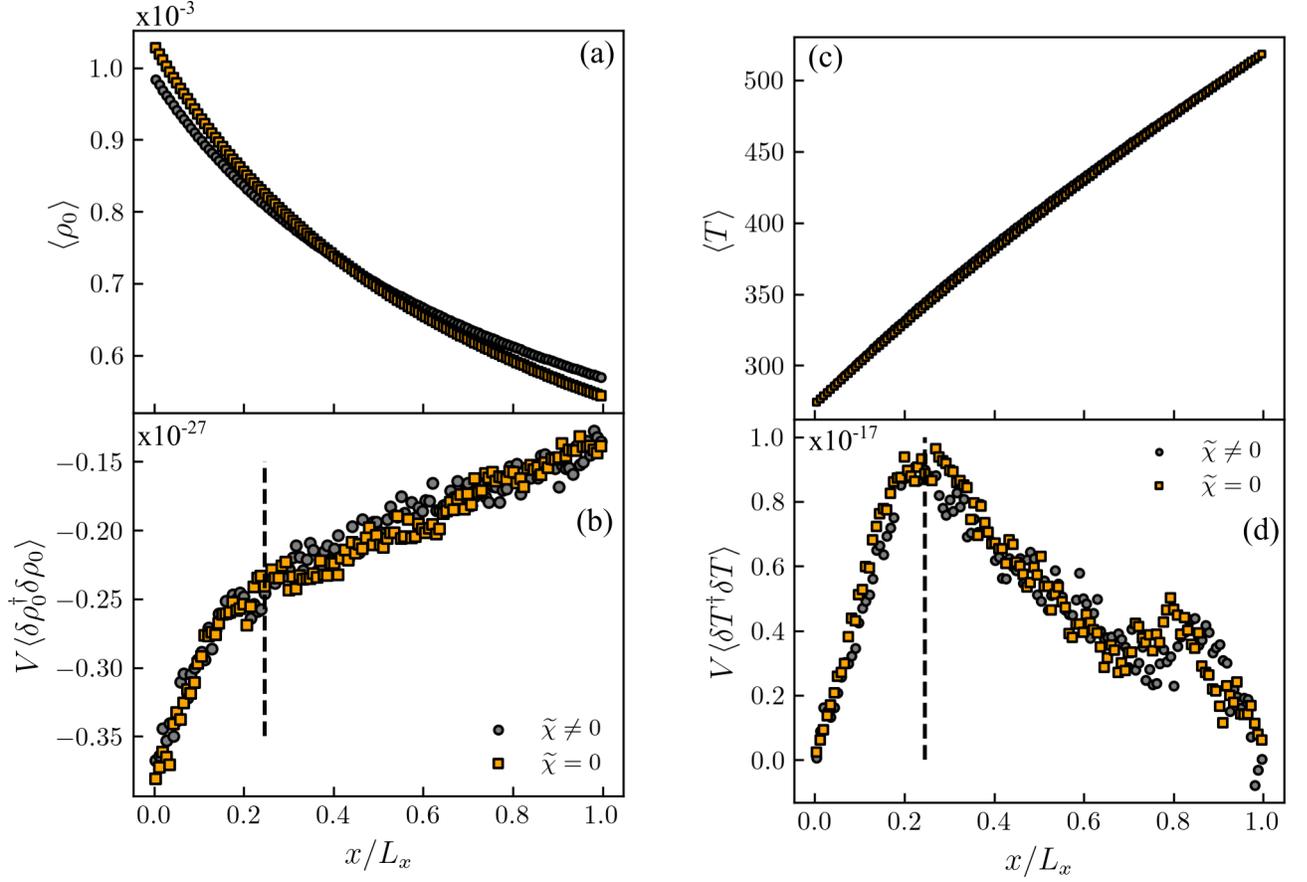

FIG. S6. (a) and (c) Mean profiles of $\langle \rho_0 \rangle$ and $\langle T \rangle$ for the two cases of $\widetilde{\chi} \neq 0$ (circles) and $\widetilde{\chi} = 0$. (b) and (d) Real-space correlations $\langle \delta\rho_0^\dagger \delta\rho_0 \rangle$ and $\langle \delta T^\dagger \delta T \rangle$ respectively for $\widetilde{\chi} \neq 0$ (circles) and $\widetilde{\chi} = 0$. The reader is referred to the caption of Fig. 8 in the main text for details.


\fi

\end{document}